\def\beq#1{\begin{equation}\label{#1}}
\def\eeq{\end{equation}}
\def\beqa#1{\begin{eqnarray}\label{#1}}
\def\eeqa{\end{eqnarray}}

\def\fun#1#2{\lower3.6pt\vbox{\baselineskip0pt\lineskip.9pt
        \ialign{$\mathsurround=0pt#1\hfill##\hfil$\crcr#2\crcr\sim\crcr}}}

\def\xi{{{\bf x}^b}}

\documentclass
[twocolumn,aps,showpacs,showkeys,nofootinbib,superscriptaddress]{revtex4}
\usepackage{epsfig}
\usepackage{graphicx}
\usepackage{dcolumn}
\usepackage{bm}
\usepackage{hyperref}

\newcommand{\be}{\begin{equation}}
\newcommand{\ee}{\end{equation}}
\newcommand{\ba}{\begin{eqnarray}}
\newcommand{\ea}{\end{eqnarray}}

\begin{document}
\input{epsf.sty}

\title{Cosmological implications of different baryon acoustic oscillation data}

\author{Shuang Wang}
\email{wangshuang@mail.sysu.edu.cn}
\affiliation{School of Physics and Astronomy, Sun Yat-Sen University, Zhuhai 519082, P. R. China}%

\author{Yazhou Hu}
 \affiliation{Kavli Institute of Theoretical Physics China, Chinese Academy of Scienses, Beijing 100190, P. R. China}

\author{Miao Li}
\affiliation{School of Physics and Astronomy, Sun Yat-Sen University, Zhuhai 519082, P. R. China}%

\begin{abstract}
In this work, we explore the cosmological implications of different baryon acoustic oscillation (BAO) data,
including the BAO data extracted by using the spherically averaged one-dimensional galaxy clustering (GC) statistics (hereafter BAO1) and
the BAO data obtained by using the anisotropic two-dimensional GC statistics (hereafter BAO2).
To make a comparison, we also take into account the case without BAO data (hereafter NO BAO).
Firstly, making use of these BAO data,
as well as the SNLS3 type Ia supernovae sample and the Planck distance priors data,
we give the cosmological constraints of the $\Lambda$CDM, the $w$CDM, and the Chevallier-Polarski-Linder (CPL) model.
Then, we discuss the impacts of different BAO data on cosmological consquences, including its effects on parameter space, equation of state (EoS), figure of merit (FoM), deceleration-acceleration transition redshift, Hubble parameter $H(z)$, deceleration parameter $q(z)$,
statefinder hierarchy $S^{(1)}_3(z)$, $S^{(1)}_4(z)$ and cosmic age $t(z)$.
We find that: (1) NO BAO data always give a smallest fractional matter density $\Omega_{m0}$, a largest fractional curvature density $\Omega_{k0}$ and a largest Hubble constant $h$; in contrast, BAO1 data always give a largest $\Omega_{m0}$, a smallest $\Omega_{k0}$ and a smallest $h$.
(2) For the $w$CDM and the CPL model, NO BAO data always give a largest EoS $w$; in contrast, BAO2 data always give a smallest $w$.
(3) Compared with the case of BAO1, BAO2 data always give a slightly larger FoM, and thus can give a cosmological constraint with a slightly
better accuracy.
(4) The impacts of different BAO data on the cosmic evolution and the comic age are very small, and can not be distinguished by using various dark energy diagnosis and the cosmic age data.

\end{abstract}


\maketitle

\section{Introduction}

Since its discovery in 1998~\citep{Riess1998}, cosmic acceleration has become one of the central problems in theoretical physics and modern cosmology.
So far, we are still in the dark about the nature of this extremely counterintuitive phenomenon; it may be due to an unknown energy component, i.e., dark energy (DE), or a modification of general relativity, i.e., modified gravity (MG) \citep{Caldwell2009}.

One of the most powerful probes of DE is baryon acoustic oscillation (BAO), which can be used as a cosmological standard ruler to measure the expansion history of the Universe \citep{Blake2003}. The BAO scale can be measured in the power spectrum of cosmic microwave background (CMB) and in the maps of large-scale structure at lower redshifts. Moreover, for the latter, the BAO data can be extracted using either correlation function analysis \citep{Eisenstein2005} or power spectrum analysis \citep{Percival2010}.

There are mainly two approaches to extract BAO scale from the galaxy redshift survey (GRS) data ~\citep{Eisenstein2005, Tegmark2006, Percival2010, Padmanabhan2012, Anderson2012, Aubourg2014}.
The first approach is adopting the spherically averaged one-dimensional (1D) galaxy clustering (GC) statistics.
For example, the BAO position in spherically averaged two point correlation functions (2PCF) provides a measure of an effective distance
$D_{\rm V}(z) \equiv [(1+z)^2{D_A(z)}^2\frac{cz}{H(z)}]^{1/3}$ (Here $z$ denotes the redshift and $c$ denotes the speed of light),
which was introduced by Eisenstein {et~al.}~\citep{Eisenstein2005} according to the different dilation scales for the Hubble parameter $H(z)$ and the angular diameter distance $D_{\rm A}(z)$.
Making using of the Sloan Digital Sky Survey Data Release 7 (SDSS DR7) \citep{Abazajian2009},
Padmanabhan {et~al.} ~\citep{Padmanabhan2012} gave $D_V(z=0.35)/r_s(z_d) = 8.88\pm0.17$;
based on the Baryon Oscillation Spectroscopic Survey Data Release 9 (BOSS DR9) \citep{Eisenstein2011},
Anderson {et~al.} \citep{Anderson2012} obtained $D_V(z=0.57)/r_s(z_d) = 13.67\pm0.22$.
Here $r_s(z)$ is the comoving sound horizon, and $z_d$ is the redshift of ``drag'' epoch when the baryons are ``released'' from the drag of the photons.

Another approach is making use of the anisotropic two-dimensional (2D) GC statistics.
The key idea is separating the line of sight and transverse clustering so as to measure $H(z)$ and $D_{\rm A}(z)$ separately~\citep{Okumura2008, anderson2014}.
In a series of works ~\citep{ChuangWang}, Chuang and Wang presented a method to obtain robust measurements of $H(z)$ and $D_A(z)$ simultaneously from the full 2D 2PCF.
Applying this method to the BOSS DR9 data, Wang gave $H(z=0.57)r_s(z_d)/c = 0.0444 \pm 0.0019$ and $D_A(z=0.57)/r_s(z_d) = 9.01 \pm 0.23$.
In addition, Hemantha, Wang and Chuang \citep{Hemantha2014} also proposed a method to measure $H(z)$ and $D_A(z)$ simultaneously from the 2D matter power spectrum (MPS);
applying this method to the SDSS DR7 sample, they obtained $H(z=0.35)r_s(z_d)/c = 0.0431 \pm 0.0018$ and $D_A(z=0.35)/r_s(z_d) = 6.48 \pm 0.25$.

Thus we have two types of BAO data: one is obtained by using the spherically averaged 1D GC statistics, another is obtained by using the anisotropic 2D GC statistics.
An important difference between these two kinds of analyses is that the anisotropic analysis contains an Alcock-Paczynksi test~\citep{APtest1979}.
Although both these two types of BAO data are widely used in the literature to test various cosmological models~\citep{Addison2013,Aubourg2014}, so far as we know,
the effects of different BAO data on cosmology-fits and corresponding cosmological consequences have not been studied in the past.
So the main aim of our work is presenting a comprehensive and systematic investigation
on the cosmological implications of different BAO data.
To make a comparison, we also take into account the case without any BAO data.

In this work, making use of these BAO data, as well as the SNLS3 type Ia supernovae (SNe Ia) data~\citep{Conley2011} and the Planck distance prior data~\citep{WangyunWangshuang2013}, we constrain the parameter spaces of three simplest DE models, including the $\Lambda$-cold-dark-matter ($\Lambda$CDM) model, the $w$CDM model, and the Chevallier-Polarski-Linder (CPL) model \citep{Chevallier}.
Moreover, based on the fitting results,
we study the impacts of different BAO data on cosmological consquences,
including its effects on parameter space, equation of state (EoS) \citep{MIDE1,MIDE2}, figure of merit (FoM)\citep{DETF2006, AB2007},
deceleration-acceleration transition redshift, Hubble parameter $H(z)$, deceleration parameter $q(z)$,
statefinder hierarchy $S^{(1)}_3(z)$ and $S^{(1)}_4(z)$~\citep{ArabSahni2011}, and cosmic age $t(z)$~\citep{Alcaniz99}.

We present our method in Section~\ref{sec:method}, our results in Section~\ref{sec:results}, and
summarize and conclude in Section~\ref{sec:conclusion}.

\section{Methodology}
\label{sec:method}

In this section, firstly we review the theoretical framework of the DE models; then we describe the observational data used in the present work; finally we introduce the background knowledge about FoM, various DE diagnosis tools and cosmic age.

\subsection{Theoretical Models}
\label{subsec:theoretical models}

In a non-flat Universe, the Friedmann equation is
\be\label{F.e.}
    3M_{pl}^{2}H^{2}=\rho_{r}+\rho_{m}+\rho_{k}+\rho_{de},
\ee
where $H \equiv \dot{a}/a$ is the Hubble parameter,
$a=(1+z)^{-1}$ is the scale factor of the Universe (we take today's scale factor $a_0=1$),
the dot denotes the derivative with respect to cosmic time $t$,
$M^2_{pl} = (8\pi G)^{-1}$ is the reduced Planck mass, $G$ is Newtonian gravitational constant,
$\rho_{r}$, $\rho_{m}$, $\rho_{k}$ and $\rho_{de}$
are the energy densities of radiation, matter, spatial curvature and DE, respectively.
The reduced Hubble parameter $E(z)\equiv H(z)/H_{0}$ satisfies
\be\label{E}
E^{2}=\Omega_{r0}(1+z)^{4}+\Omega_{m0}(1+z)^{3}+\Omega_{k0}(1+z)^2+\Omega_{de0}f(z),
\ee
where $H_0$ is the Hubble constant, $\Omega_{r0}$, $\Omega_{m0}$, $\Omega_{k0}$ and $\Omega_{de0}$
are the present fractional densities of radiation, matter, spatial curvature and DE, respectively.
Per \citep{WangWang2013}, we take $\Omega_{r0}=\Omega_{m0} / (1+z_{\rm eq})$, where $z_{\rm eq}=2.5\times 10^4 \Omega_{m0} h^2 (T_{\rm cmb}/2.7\,{\rm K})^{-4}$ and $T_{\rm cmb}=2.7255\,{\rm K}$.
Since $\Omega_{de0}=1-\Omega_{m0}-\Omega_{r0}-\Omega_{k0}$, $\Omega_{de0}$ is not an independent parameter \citep{Zhangetal2012}.
Here the DE density function $f(z) \equiv \rho_{de}(z)/\rho_{de}(0)$, which satisfies
\be\label{Density function.e.}
    f(z)={\rm exp}[3\int_{0}^{z}dz^{\prime}\frac{1+w(z^{\prime})}{1+z^{\prime}}],
\ee
where the EoS $w$ is the ratio of pressure to density for the DE
\be
\label{eq:eos}
w = p_{de}/\rho_{de}
\ee

In the present work we just consider three simplest DE models:
\begin{itemize}
\item
$\Lambda$CDM model, which has a cosmological constant (i.e. $w = -1$).
Then we have
\ba\label{E_lcdm}
E(z)=\Big(\Omega_{r0}(1+z)^{4}+\Omega_{m0}(1+z)^{3}+\Omega_{k0}(1+z)^2
\nonumber\\
+\Omega_{de0}\Big)^{1/2},
\ea
\item
$w$CDM model, which has a constant $w$.
Then we have
\ba\label{E_wcdm}
E(z)=\Big(\Omega_{r0}(1+z)^{4}+\Omega_{m0}(1+z)^{3}+\Omega_{k0}(1+z)^2
\nonumber\\
+\Omega_{de0}(1+z)^{3(1+w)}\Big)^{1/2},
\ea
\item
CPL model \citep{Chevallier} has a dynamical $w(z) = w_{0} + w_{a}z/(1+z)$.
Then we have
\ba\label{E_cpl}
E(z)=\Big(\Omega_{r0}(1+z)^{4}+\Omega_{m0}(1+z)^{3}+\Omega_{k0}(1+z)^2
\nonumber\\
+\Omega_{de0}(1+z)^{3(1+w_0+w_a)}e^{\frac{-3w_az}{1+z}}\Big)^{1/2},
\ea
\end{itemize}
For each model, the expression of $E(z)$ will be used to calculate the observational quantities appearing in the next subsection.

\subsection{Observational Data}

In this subsection, we describe the observational data used in this work.

\subsubsection{BAO Data}

In this work, we make use of two types of BAO data extracted from SDSS DR7 \citep{Abazajian2009} and the BOSS DR9 \citep{Eisenstein2011}.

\begin{itemize}
\item
Let us introduce the BAO data obtained by using the spherically averaged 1D galaxy clustering statistics first.
As mentioned above,
making using of the SDSS DR7,
Padmanabhan {et~al.} ~\citep{Padmanabhan2012} gave
\be
D_V(z=0.35)/r_s(z_d) = 8.88\pm0.17;
\ee
based on the BOSS DR9,
Anderson {et~al.} \citep{Anderson2012} obtained
\be
D_V(z=0.57)/r_s(z_d) = 13.67\pm0.22.
\ee

The effective distance $D_V(z)$ is given by ~\citep{Eisenstein2005},
\be
D_V(z) \equiv [(1+z)^2{D_A(z)}^2\frac{cz}{H(z)}]^{1/3},
\ee
where the angular diameter distance
\be
D_{A}(z) = c H_0^{-1}r(z)/(1+z),
\ee
and the comoving distance
\be
\label{eq:rz}
r(z)=cH_0^{-1}\, |\Omega_k|^{-1/2} {\rm sinn}[|\Omega_k|^{1/2}\, \Gamma(z)].
\ee
Here $\Gamma(z)=\int_0^z\frac{dz'}{E(z')}$,
${\rm sinn}(x)=\sin(x)$, $x$, $\sinh(x)$ for $\Omega_k<0$, $\Omega_k=0$, and $\Omega_k>0$, respectively.
In addition, the comoving sound horizon $r_s(z)$ is given by \citep{WangWang2013}
\be~\label{sound_horizon}
r_s(z) = cH_0^{-1}\int_{0}^{a}\frac{da^{\prime}}{\sqrt{3(1+\overline{R_b}a^\prime){a^\prime}^4E^2(z^\prime)}},
\ee
where $\overline{R_b}=31500\Omega_bh^2(T_{CMB}/2.7K)^{-4}$,
and $\Omega_b$ is the present fractional density of baryon.
The redshift of the drag epoch $z_d$ is well approximated by \cite{EisensteinHu1998}
\be
z_d  =
 \frac{1291(\Omega_mh^2)^{0.251}}{1+0.659(\Omega_mh^2)^{0.828}}
\left[1+b_1(\Omega_bh^2)^{b2}\right],
\label{eq:zd}
\ee
where
\ba
  b_1 &= &0.313(\Omega_mh^2)^{-0.419}\left[1+0.607(\Omega_mh^2)^{0.674}\right],  \nonumber \\
  b_2 &= &0.238(\Omega_mh^2)^{0.223}.
\ea

These BAO data are included in our analysis by adding $\chi^2_{BAO}=\chi^2_{1}+\chi^2_{2}$,
with $z_{1}=0.35$ and $z_{2}=0.57$, to the $\chi^2$ of a given model.
Then we have
\be
\label{eq:chi2bao1}
\chi^2_{i}=\Big(\frac{q_i - q_i^{data}}{\sigma_i}\Big)^2
\ee
where $\sigma_i$ is the standard deviation of data, $q_i=D_{\rm V}(z_i)/r_{\rm s}(z_d)$, and $i=1,2$.
For simplicity, hereafter we will call this type BAO data ``BAO1''.

\item
Then, let us introduce the BAO data obtained by using the anisotropic 2D galaxy clustering statistics.
Making use of the 2D MPS of SDSS DR7 samples,
Hemantha, Wang, and Chuang~\citep{Hemantha2014} got
\ba
H(z=0.35)r_s(z_d)/c&=&0.0431  \pm  0.0018,  \nonumber \\
D_A(z=0.35)/r_s(z_d)&=& 6.48  \pm  0.25.
\label{eq:CW2}
\ea
In addition, using the 2D 2PCF of BOSS DR9 samples,
Wang~\cite{Wangyun2014} obtained
\ba
H(z=0.57)r_s(z_d)/c&=&0.0444	\pm  0.0019,  \nonumber \\
D_A(z=0.57)/r_s(z_d)&=& 9.01	\pm  0.23.
\label{eq:C13}
\ea
These BAO data are included in our analysis by adding $\chi^2_{BAO}=\chi^2_{1}+\chi^2_{2}$,
with $z_{1}=0.35$ and $z_{2}=0.57$, to the $\chi^2$ of a given model.
Now we have
\be
\label{eq:chi2bao2}
\chi^2_{i}=\Delta q_i \left[ {\rm C}^{-1}_{i}(q_i,q_j)\right]
\Delta q_j,
\hskip .2cm
\Delta q_i= q_i - q_i^{data},
\ee
where $q_1=H(z_{i})r_s(z_d)/c$, $q_2=D_A(z_{i})/r_s(z_d)$, and $i=1,2$.
Based on Refs.~\citep{Hemantha2014} and~\citep{Wangyun2014}, we get
\begin{equation}
 {\rm C}_{1}=\left(
  \begin{array}{cc}
    0.00000324 & -0.00010728 \\
    -0.00010728 & 0.0625 \\
  \end{array}
\right),
\end{equation}
\begin{equation}
 {\rm C}_{2}=\left(
  \begin{array}{cc}
    0.00000361 & 0.0000176111 \\
    0.0000176111 & 0.0529 \\
  \end{array}
\right).
\end{equation}
For simplicity, hereafter we will call this type BAO data ``BAO2''.
\end{itemize}

\subsubsection{SNe Ia Data}

For the SNe Ia data, we use the SNLS3 ``combined'' sample~\citep{Conley2011}, which consisting of 472 SNe Ia.
The $\chi^2$ function for the supernova (SN) data is given by
\begin{equation}\label{SNchisq}
\chi^2_{SNLS3}=\Delta \overrightarrow{\bf m}^T \cdot {\bf C}^{-1} \cdot \Delta \overrightarrow{\bf m},
\end{equation}
where $\Delta {\overrightarrow {\bf m}} = {\overrightarrow {\bf m}}_B - {\overrightarrow {\bf m}}_{\rm mod}$ is a vector of
model residuals of the SN sample,
and $m_B$ is the rest-frame peak $B$ band magnitude of the SN.
The total covariance matrix $\mbox{\bf C}$ can be written as~\citep{Conley2011}
\be
\mbox{\bf C}=\mbox{\bf D}_{\rm stat}+\mbox{\bf C}_{\rm stat}+\mbox{\bf C}_{\rm sys}.
\ee
Here $\mbox{\bf D}_{\rm stat}$ denotes the diagonal part of the statistical uncertainty,
$\mbox{\bf C}_{\rm stat}$ and $\mbox{\bf C}_{\rm sys}$ denote the statistical and systematic covariance matrices, respectively.
For the details of constructing the covariance matrix $\mbox{\bf C}$, see \cite{Conley2011}.

Current studies have found the evidence for the potential SN evolution.
For examples, the studies on the SNLS3~\citep{WangWang2013}, the Union2.1~\citep{Mohlabeng2013}, the Pan-STARRS1~\citep{Scolnic2014}, and the JLA data sets~\citep{jlapotential}
all indicated that, although the SN stretch-luminosity parameter $\alpha$ is still consistent with a constant, the SN color-luminosity parameter $\beta$ evolves along with redshift $z$ at very high confidence level (CL).
Moreover, this conclusion has significant effects on parameter estimation of various cosmological models~\citep{WLZ2014}.
Therefore, in the present work we adopt a constant $\alpha$ and a linear $\beta = \beta_0 + \beta_1 z$ , now the predicted magnitude of SN becomes
\footnote{It should be mentioned that, the intrinsic scatter $\sigma_{\rm int}$ also has the hint of redshift-dependence~\citep{Marrinerl2011}
that will significantly affect the results of cosmology-fits. In addition, different light-curve fitters of SN can also affect the results of parameter estimation~\citep{Bengochea2011,HLLW2015}.
But in this work, we do not consider these factors for simplicity.}
\be~\label{snvec_betaz}
m_{\rm mod}=5 \log_{10}{\cal D}_L(z) - \alpha (s-1) +\beta(z) {\cal C} + {\cal M}.
\ee
The luminosity distance ${\cal D}_L(z)$ is defined as
\be
{\cal D}_L(z)\equiv H_0 (1+z_{\rm hel}) r(z),
\ee
where $z$ and $z_{\rm hel}$ are the CMB restframe and heliocentric redshifts of SN,
and $r(z)$ has been given in Eq. (\ref{eq:rz}).
In addition, $s$ and ${\cal C}$ are stretch measure and color measure for the SN light curve,
$\mathcal{M}$ is a parameter representing some combination of the absolute magnitude $M$ of a fiducial SNe Ia and the Hubble constant $H_0$.

It must be emphasized that, in order to include host-galaxy information in the cosmological fits,
Conley et al. \citep{Conley2011} split the SNLS3 sample based on host-galaxy stellar mass at $10^{10} M_{\odot}$,
and made ${\cal M}$ to be different for the two samples.
So there are two values of ${\cal M}$ (i.e. ${\cal M}_1$ and ${\cal M}_2$) for the SNLS3 data.
Moreover, Conley et al. removed ${\cal M}_1$ and ${\cal M}_2$ from cosmology-fits by analytically marginalizing over them
(for more details, see the appendix C of~\citep{Conley2011}).
In the present work, we just follow the recipe of~\citep{Conley2011}, and do not treat ${\cal M}$ as model parameter.
For simplicity, hereafter we will call this SNe Ia sample ``SNLS3''.

\subsubsection{CMB Data}

For cosmic microwave background (CMB) data, we use the Planck distance priors data extracted from Planck first data release~\citep{WangyunWangshuang2013}.
CMB gives us the comoving distance to the photon-decoupling surface $r(z_*)$ and the comoving sound horizon at photon-decoupling epoch $r_s(z_*)$.
Wang and Mukherjee \cite{WangMukherjee2007} showed that the CMB shift parameters
\ba
l_a &\equiv &\pi r(z_*)/r_s(z_*), \nonumber\\
R &\equiv &\sqrt{\Omega_m H_0^2} \,r(z_*)/c,
\ea
together with $\omega_b\equiv \Omega_b h^2$, provide an efficient summary
of CMB data as far as dark energy constraints go.
Notice that $r(z)$ is given in Eq. (\ref{eq:rz}), $r_s(z)$ is given in Eq. \ref{sound_horizon}, and $z_*$ is given in Ref. \citep{HuSugiyama1996}.
In \citep{Li2008}, Li et al. proved that CMB distance priors data can give similar constraints on DE parameters compared with the full CMB power spectrum data.
This conclusion is also consistent with the result of \citep{LiYH13}, in which the holographic DE model is adopted in the background.
Therefore, the use of the Planck distance prior is sufficient for our purpose.

Using the Planck+lensing+WP data,
the mean values and 1$\sigma$ errors of $\{ l_a, R, \omega_b\}$ are obtained ~\citep{WangyunWangshuang2013},
\ba
&&\langle l_a \rangle = 301.57, \sigma(l_a)=0.18, \nonumber\\
&&\langle R \rangle = 1.7407,  \sigma(R)=0.0094, \nonumber\\
&& \langle \omega_b \rangle = 0.02228, \sigma(\omega_b)=0.00030.
\label{eq:CMB_mean_planck}
\ea
Defining $p_1=l_a(z_*)$, $p_2=R(z_*)$, and $p_3=\omega_b$,
the normalized covariance matrix $\mbox{NormCov}_{CMB}(p_i,p_j)$ can be written as~\citep{WangyunWangshuang2013}
\be
\left(
\begin{array}{ccc}
  1.0000  &    0.5250  &   -0.4235    \\
  0.5250  &    1.0000  &   -0.6925    \\
 -0.4235  &   -0.6925  &    1.0000    \\
\end{array}
\right).
\label{eq:normcov_planck}
\ee
Then, the covariance matrix for $(l_a, R, \omega_b)$ is given by
\be
\mbox{Cov}_{CMB}(p_i,p_j)=\sigma(p_i)\, \sigma(p_j) \,\mbox{NormCov}_{CMB}(p_i,p_j),
\label{eq:CMB_cov}
\ee
where $i,j=1,2,3$.
The CMB data are included in our analysis by adding
the following term to the $\chi^2$ function:
\be
\label{eq:chi2CMB}
\chi^2_{CMB}=\Delta p_i \left[ \mbox{Cov}^{-1}_{CMB}(p_i,p_j)\right]
\Delta p_j,
\hskip .2cm
\Delta p_i= p_i - p_i^{data},
\ee
where $p_i^{data}$ are the mean values from Eq. (\ref{eq:CMB_mean_planck}).
For simplicity, hereafter we will call this CMB data ``Planck''.

\subsubsection{Total $\chi^2$ function}

Now the total $\chi^2$ function is
\be
\chi^2=\chi^2_{BAO}+\chi^2_{SNLS3}+\chi^2_{Planck}.
\ee
We perform an MCMC likelihood analysis using the ``CosmoMC'' package~\citep{Lewis2002}.

\subsection{Figure of Merit, Dark Energy Diagnosis and Cosmic Age}

Let us start from the FoM of DE.
FoM was firstly proposed to make a comparison for different DE experiments.
Making use of the CPL model, Dark Energy Task Force (DETF) developed a quantitive FoM to be the reciprocal of the area enclosed by the $95\%$ CL contour in the $(w_0, w_a)$ plane, satisfies \citep{DETF2006}
\be\label{fomr}
\rm~FoM_{DETF}=\frac{1}{6.17\pi\sigma(w_a)\sigma(w_p)}
\ee
where $\sigma(w_p)=w_0-w_a{\langle\delta w_0\delta w_a\rangle/}\langle{\delta w_a}^2\rangle$, and $\sigma(w_i)=\sqrt{\langle{\delta w_i}^2\rangle}$.
Note that $\sigma(w_a)\sigma(w_p)=\sqrt{\rm~{detCov}(w_0, w_a)}$, thus the conversion to $w_p$ is not needed to calculate the FoM.
Soon after, various FoM quantities were proposed \citep{AB2007}.
In this work, we just use the DETF version of FoM to make a comparison between different BAO data.
It is clear that a larger $\rm~FoM$ indicates a better accuracy.

Then let us turn to various DE diagnosis.
The scale factor of the Universe $a$ can be Taylor expanded around today's cosmic age $t_0$ as follows:
\be
a(t)=1+\sum\limits_{\emph{n}=1}^{\infty}\frac{A_{\emph{n}}}{n!}[H_0(t-t_0)]^n,
\ee
where
\be
A_{\emph{n}}=\frac{a(t)^{(n)}}{a(t)H^n},~~n\in N,
\ee
with $a(t)^{(n)}=d^na(t)/dt^n$.
The Hubble parameter $H(z)$ contains the information of the first derivative of $a(t)$.
Based on the BAO measurements from the BOSS DR9 and DR 11,
Samushia et al.~\citep{Samushia2013} gave $H_{0.57}\equiv H(z=0.57)=92.4\pm4.5 {\rm km/s/Mpc}$,
while Delubac et al.~\citep{Delubac2015} obtained $H_{2.34}\equiv H(z=2.34)=222\pm7 {\rm km/s/Mpc}$.
These two H(z) data points will be used to compare the theoretical predictions of DE models.

In addition, the deceleration parameter $q$ is
\be
q=-A_2=-\frac{\ddot{a}}{aH^{2}},
\ee
which contains the information of the second derivatives of $a(t)$.
It is clear that the expansion of the Universe underwent a transition from the deceleration phase to the acceleration phase in the past.
The deceleration-acceleration transition redshift $z_t$ can be calculate using the condition
\be
q(z_t)=0.
\ee
Lima {et~al.}~\citep{Lima2012} argued that $z_t$ may be tightly constrained by ongoing and future observations.
So it is interesting to check the impacts of different BAO data on the transition redshift $z_t$.

The statefinder hierarchy, $S_{\emph{n}}$, is defined as~\citep{ArabSahni2011}:
\ba
&S_{2}=A_{2}+\frac{3}{2}\Omega_{\rm m},\\
&S_{3}=A_{3},\\
&S_{4}=A_{4}+\frac{3^2}{2}\Omega_{\rm m},
\ea
The reason for this redefinition is to peg the statefinder at unity for $\Lambda$CDM during the cosmic expansion,
\be
S_{\emph{n}}|_{\Lambda \rm{CDM}}=1.
\ee
This equation defines a series of null diagnostics for $\Lambda$CDM when $n\geq3$.
By using this diagnostic, we can easily distinguish the $\Lambda$CDM model from other DE models.
Because of $\Omega_{m}|_{\Lambda \rm{CDM}}=\frac{2}{3}(1+q)$, when $n\geq3$, statefinder hierarchy can be rewritten as:
\ba
&S^{(1)}_{3}=A_{3},\\
&S^{(1)}_{4}=A_{4}+3(1+q),
\ea
where the superscript $(1)$ is to discriminate between $S^{(1)}_{\emph{n}}$ and $S_{\emph{n}}$.
In this work, we use the statefinder hierarchy $S^{(1)}_3(z)$ and $S^{(1)}_4(z)$ to diagnose the impacts of different types BAO data on the three DE models.

At last, we introduce the cosmic age.
The age of the Universe at redshift $z$ is given by
\be
t(z) =\int_z^\infty\frac{d\tilde{z}}{(1+\tilde{z})H(\tilde{z})}.
\ee
In history, the cosmic age problem played an important role in the cosmology \cite{Alcaniz99}.
Obviously, the age of the Universe at any redshift $z$ cannot be younger than its constituents at the same redshift.
In the literature, some old high redshift objects (OHRO) have been considered extensively.
For instance,
the 3.5Gyr old galaxy LBDS 53W091 at redshift $z = 1.55$ \cite{Dunlop96},
the 4.0Gyr old galaxy LBDS 53W069 at redshift $z = 1.43$ \cite{Dunlop99},
and the 2.0Gyr old quasar APM 08279+5255 at redshift $z = 3.91$ \cite{Hasinger02}.
The age data of these three OHRO
(i.e. $t_{1.43} \equiv t(z=1.43)=4.0 Gyr$, $t_{1.55} \equiv t(z=1.55)=3.5 Gyr$ and $t_{3.91} \equiv t(z=3.91)=2.0 Gyr$)
have been extensively used to test various cosmological models in the literature (see e.g. \cite{Alcaniz03}).

\section{Results}
\label{sec:results}

\subsection{Cosmology Fits and Corresponding Cosmological Consequences}

In this subsection, we present the fitting results of the three DE models,
and discuss the impacts of different BAO data on parameter estimation and FoM.
For comparison, we also take into account the case without using BAO data.
Hereafter, ``NO BAO'', ``BAO1'' and ``BAO2'' represent the $\rm SNLS3+Planck$, the $\rm BAO1+SNLS3+Planck$ and the $\rm BAO2+SNLS3+Planck$ data, respectively.

\begin{table*}\scriptsize
\centering
\caption{Cosmological fitting results for the $\Lambda$CDM, the $w$CDM and the CPL model, where both the best-fit values and the $1\sigma$ errors of various parameters are listed.
``NO BAO'', ``BAO1'' and ``BAO2'' represent the results given by the SNLS3+Planck, the BAO1+SNLS3+Planck and the BAO2+SNLS3+Planck data, respectively.}
\label{tab:res_cos}
\centering
\begin{tabular}{ccccccccccccc}
\hline\hline &\multicolumn{3}{c}{$\Lambda$CDM}&&\multicolumn{3}{c}{$w$CDM}&&\multicolumn{3}{c}{CPL} \\
           \cline{2-4}\cline{6-8}\cline{10-12}
           Parameter  & NO BAO & BAO1 & BAO2 & & NO BAO & BAO1 & BAO2 & & NO BAO & BAO1 & BAO2    \\ \hline
$\alpha$           & $1.419^{+0.071}_{-0.071}$
                   & $1.417^{+0.070}_{-0.076}$
                   & $1.418^{+0.071}_{-0.072}$&
                   & $1.446^{+0.096}_{-0.112}$
                   & $1.419^{+0.071}_{-0.071}$
                   & $1.420^{+0.070}_{-0.077}$&
                   & $1.441^{+0.028}_{-0.093}$
                   & $1.418^{+0.072}_{-0.072}$
                   & $1.417^{+0.070}_{-0.077}$\\ 

$\beta_0$          & $1.434^{+0.267}_{-0.267}$
                   & $1.407^{+0.261}_{-0.260}$
                   & $1.424^{+0.259}_{-0.258}$&
                   & $1.526^{+0.399}_{-0.357}$
                   & $1.430^{+0.267}_{-0.267}$
                   & $1.466^{+0.264}_{-0.265}$&
                   & $1.475^{+0.216}_{-0.314}$
                   & $1.419^{+0.268}_{-0.268}$
                   & $1.417^{+0.288}_{-0.267}$\\ 

$\beta_1$          & $5.140^{+0.716}_{-0.796}$
                   & $5.206^{+0.709}_{-0.777}$
                   & $5.160^{+0.705}_{-0.772}$&
                   & $4.935^{+0.953}_{-1.150}$
                   & $5.146^{+0.724}_{-0.794}$
                   & $5.052^{+0.715}_{-0.790}$&
                   & $4.954^{+0.912}_{-0.582}$
                   & $5.182^{+0.720}_{-0.799}$
                   & $5.195^{+0.723}_{-0.801}$\\ 

$\Omega_{k0}$      & $0.0032^{+0.0109}_{-0.0088}$
                   & $-0.0005^{+0.0024}_{-0.0024}$
                   & $0.0016^{+0.0027}_{-0.0027}$&
                   & $0.0369^{+0.0401}_{-0.0311}$
                   & $-0.0010^{+0.0028}_{-0.0030}$
                   & $0.0005^{+0.0031}_{-0.0032}$&
                   & $-0.0051^{+0.0193}_{-0.0273}$
                   & $-0.0109^{+0.0027}_{-0.0035}$
                   & $-0.0100^{+0.0029}_{-0.0033}$\\

$\Omega_{b0}$      & $0.0452^{+0.0064}_{-0.0072}$
                   & $0.0478^{+0.0011}_{-0.0011}$
                   & $0.0462^{+0.0015}_{-0.0015}$&
                   & $0.0281^{+0.0140}_{-0.0199}$
                   & $0.0474^{+0.0017}_{-0.0017}$
                   & $0.0455^{+0.0018}_{-0.0018}$&
                   & $0.0425^{+0.0144}_{-0.0089}$
                   & $0.0462^{+0.0017}_{-0.0017}$
                   & $0.0447^{+0.0018}_{-0.0018}$\\ 

$\Omega_{m0}$      & $0.284^{+0.040}_{-0.045}$
                   & $0.301^{+0.007}_{-0.007}$
                   & $0.290^{+0.009}_{-0.009}$&
                   & $0.176^{+0.088}_{-0.125}$
                   & $0.298^{+0.010}_{-0.011}$
                   & $0.286^{+0.011}_{-0.011}$&
                   & $0.267^{+0.091}_{-0.054}$
                   & $0.291^{+0.011}_{-0.011}$
                   & $0.282^{+0.011}_{-0.011}$\\ 

$h$                & $0.708^{+0.045}_{-0.061}$
                   & $0.683^{+0.008}_{-0.008}$
                   & $0.694^{+0.011}_{-0.011}$&
                   & $0.891^{+0.089}_{-0.163}$
                   & $0.686^{+0.012}_{-0.012}$
                   & $0.700^{+0.013}_{-0.014}$&
                   & $0.724^{+0.088}_{-0.099}$
                   & $0.694^{+0.012}_{-0.012}$
                   & $0.706^{+0.014}_{-0.014}$\\ 

$w_0$              & $-$
                   & $-$
                   & $-$&
                   & $-0.857^{+0.248}_{-0.138}$
                   & $-1.022^{+0.059}_{-0.059}$
                   & $-1.038^{+0.056}_{-0.056}$&
                   & $-0.633^{+0.144}_{-0.189}$
                   & $-0.655^{+0.154}_{-0.154}$
                   & $-0.582^{+0.148}_{-0.148}$\\ 
$w_a$              & $-$
                   & $-$
                   & $-$&
                   & $-$
                   & $-$
                   & $-$&
                   & $-2.720^{+1.507}_{-2.393}$
                   & $-2.939^{+1.175}_{-1.176}$
                   & $-3.485^{+1.085}_{-1.090}$\\ 
\hline
\end{tabular}
\end{table*}

In table~\ref{tab:res_cos}, we list the cosmological fitting results for the $\Lambda$CDM, the $w$CDM and the CPL model.
From this table we see that, for all the three DE models, including the BAO data in the analysis can remarkbly improve the fitting results.
After making use of the BAO data, we see that for the $\Lambda$CDM and the $w$CDM model, the result of $\Omega_{k0}$ is consistent with the case of a spatially flat universe at $1\sigma$ CL,
which is consistent with the results of pervious studies ~\citep{Padmanabhan2012,Anderson2012,Sanchez2014,Planck201513}
\footnote{For example, the Planck group~\citep{Planck201513} investigated a non-flat $\Lambda$CDM model, and found that $\Omega_{k0}=0.000\pm0.005$.}; in contrast, for the CPL model, both the result of $\Omega_{k0}$ given by BAO1 and BAO2 deviates from a spatially flat universe at $3\sigma$ CL,
showing that there is a strong degeneracy between the evolution of EoS $w$ and the spatial curvature $\Omega_{k}$ \citep{CCB2007}.
Moreover, we find that among these three datasets, NO BAO data always give a smallest fractional matter density $\Omega_{m0}$, a largest fractional curvature density $\Omega_{k0}$ and a largest Hubble constant $h$; in contrast, BAO1 data always give a largest $\Omega_{m0}$, a smallest $\Omega_{k0}$ and a smallest $h$.
Since these results hold true for all the three DE models, we can conclude that our conclusion is insensitive to the DE considered in the background.

\begin{figure*}
  \centering
 \resizebox{0.72\columnwidth}{!}{\includegraphics{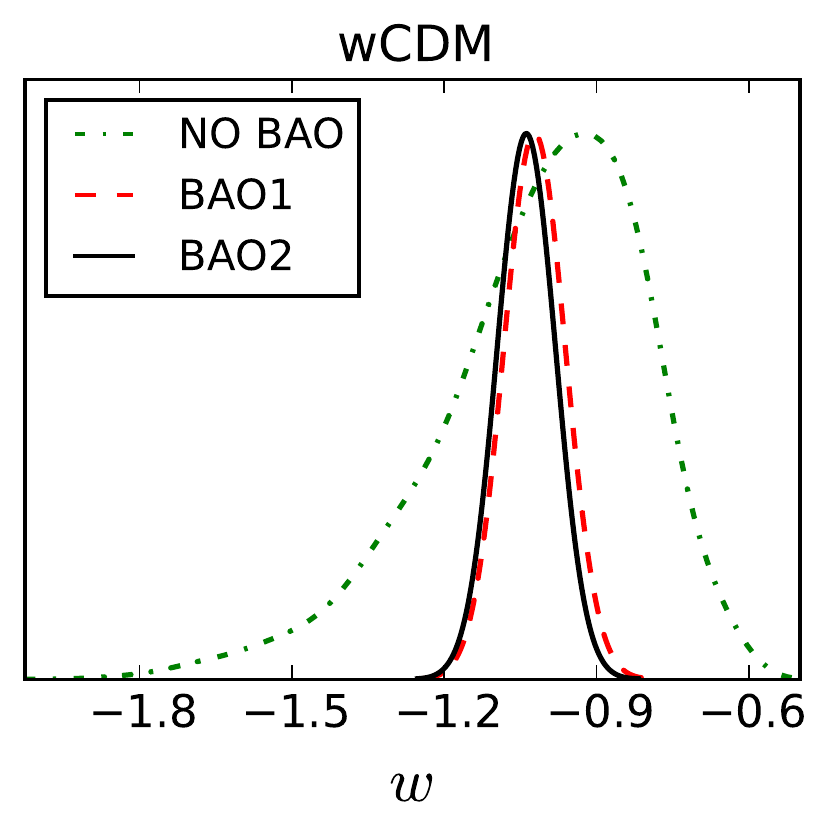}}
\caption{(color online). 1D marginalized probability distributions of $w$ for the $w$CDM model.
``NO BAO'' (green dash-dotted line), ``BAO1'' (red dashed line) and ``BAO2'' (black solid line) denote the results given by the SNLS3+Planck, the BAO1+SNLS3+Planck and the BAO2+SNLS3+Planck data, respectively.}
\label{fig:wcdm_w}
\end{figure*}

Let us discuss the issues of EoS $w$ in details.
In Fig.~\ref{fig:wcdm_w}, we plot the 1D marginalized probability distributions of $w$ for the $w$CDM model.
We see that, compared with the case without BAO data, adding BAO data in the analysis will yield a much smaller $w$;
in other words, using BAO data will yield a best-fit result $w_{bf}<-1$, while NO BAO data will lead to a best-fit result $w_{bf}>-1$.
Besides, making use of BAO data will also significantly reduce the error bar of $w$, which corresponds to a better accuracy in parameter estimation.
Moreover, we find that between these two types of BAO data, BAO2 data gives a slightly smaller $w$, as well as a slightly smaller error bar of $w$.

\begin{figure*}
  \centering
  \resizebox{0.72\columnwidth}{!}{\includegraphics{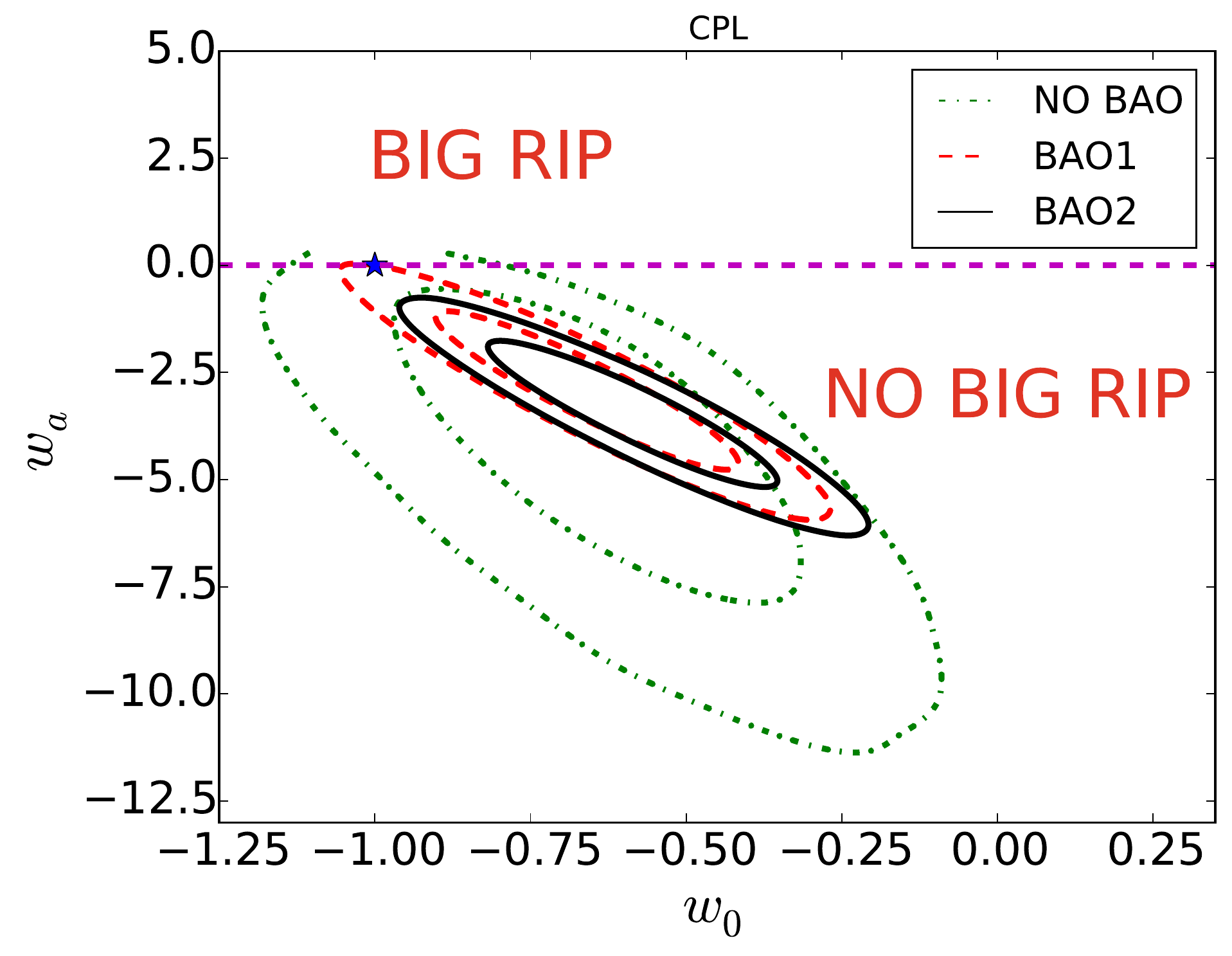}}
  \hspace{0.1\columnwidth}
  \resizebox{0.78\columnwidth}{!}{\includegraphics{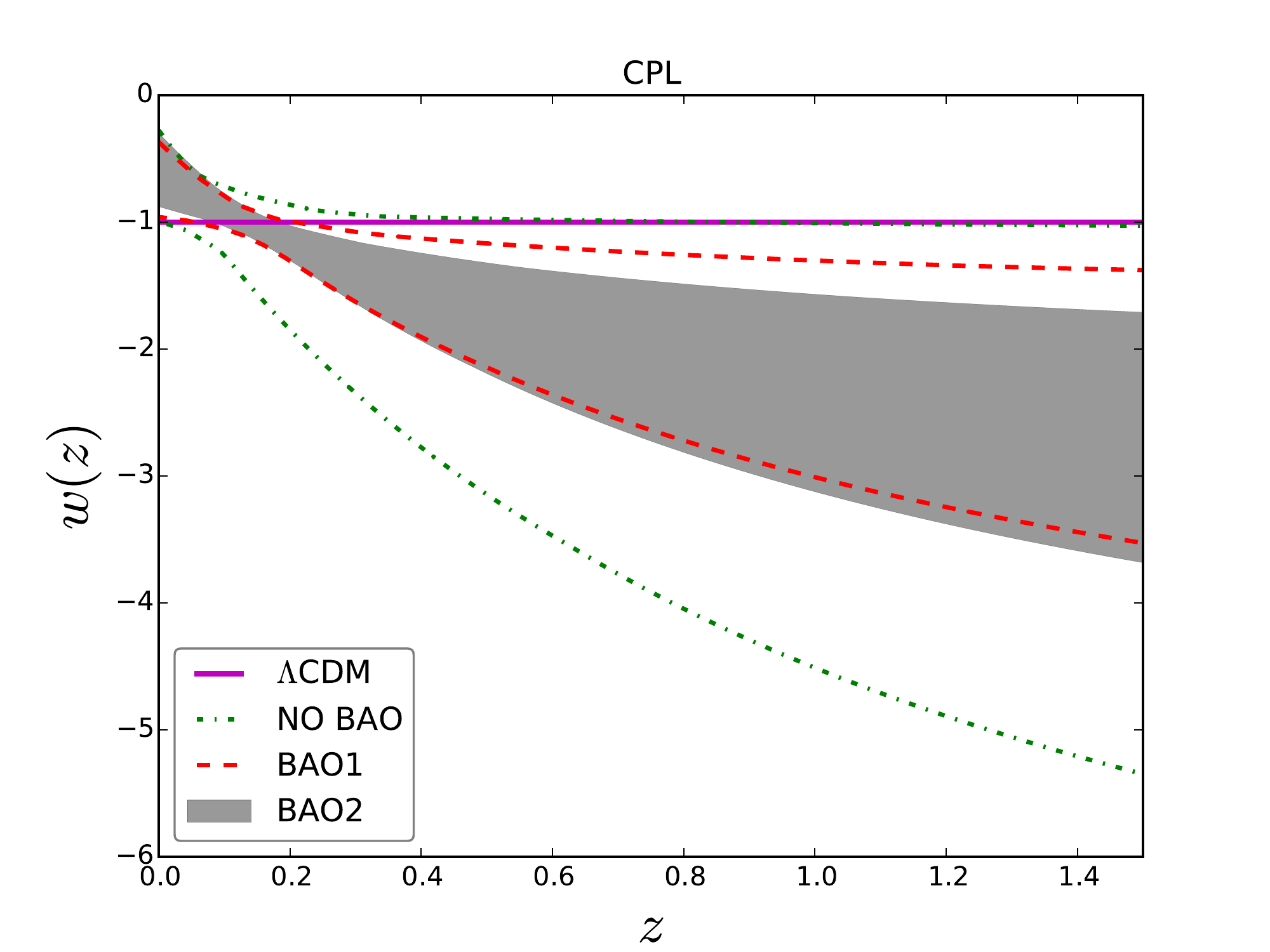}}
  \hspace{0.1\columnwidth}
  \caption{(color online). A detailed analysis on the EoS $w(z)$ of the CPL model,
including the 2D probability contours of $\{w_0, w_a\}$ at 1$\sigma$ and 2$\sigma$ CL (left panel) and
the $2\sigma$ confidence regions of $w(z)$ (right panel).
``NO BAO'', ``BAO1'' and ``BAO2'' denote the results given by the SNLS3+Planck, the BAO1+SNLS3+Planck and the BAO2+SNLS3+Planck data, respectively.
In the left panel, to make a comparison, the fixed point $\{w_0, w_a\} = \{-1,0\}$ for the $\Lambda$CDM model is also marked as a blue star;
the magenta dashed line divides the panel into two regions:
the region above the dividing line denotes a phantom dominated Universe (with big rip),
and the region below the dividing line represents a quintessence dominated Universe (without big rip).
In the right panel, to make a comparison, $w= -1$ for the $\Lambda$CDM model is also marked by a magenta horizontal line.
}
\label{fig:cpl_wz}
\end{figure*}

In the Fig.~\ref{fig:cpl_wz}, we give a detailed analysis on the EoS $w(z)$ of the CPL model.
The left panel of Fig.~\ref{fig:cpl_wz} shows the 2D probability contours of $\{w_0, w_a\}$ at 1$\sigma$ and 2$\sigma$ CL.
We see that, compared with the case of NO BAO, both BAO1 and BAO2 correspond to a significantly tighter 2D
contours of $\{w_0, w_a\}$, which implies that adding BAO data can significantly improve the fitting results.
In addition, the fixed point $\{w_0, w_a\}$ = $\{-1, 0\}$ of the $\Lambda$CDM model is located at the edge of
$2\sigma$ CL contour given by the BAO1 data, but lies outside the $2\sigma$ CL contour given by the BAO2 data.
This means that the result of BAO1 is closer to the $\Lambda$CDM model.
The right panel of Fig.~\ref{fig:cpl_wz} shows the $2\sigma$ confidence regions of $w(z)$ in the redshift range $0<z<1.5$.
From this figure we see that, between the two types of BAO data, BAO2 yields a slightly smaller $w(z)$.

\begin{table*}
\caption{The dark energy $\rm~FoM_{DETF}$ of the CPL model.
}
\vspace{2mm}
\label{tab:res_fomr}
\centering
\begin{tabular}{ccccc}
\hline\hline
      Data       &  NO BAO & BAO1 & BAO2   \\ \hline

$\rm~FoM_{DETF}$     & $0.139$
                     & $0.727$
                     & $0.804$\\ 
\hline
\end{tabular}
\end{table*}

Table~\ref{tab:res_fomr} gives the results of $\rm~FoM_{DETF}$ for the CPL model.
From this table we can see that, compared to the case of NO BAO, both the BAO1 and the BAO2 data give a much large value of $\rm~FoM_{DETF}$:
adding BAO1 will increase the value of $\rm~FoM_{DETF}$ by a factor of $5.23$,
while adding BAO2 will increase the value of $\rm~FoM_{DETF}$ by a factor of $5.78$.
This result shows the importance of using BAO measurements in measuring the cosmic expansion history and testing DE models.
In addition, for the two types of BAO data,
BAO2 can yield a slightly larger $\rm~FoM_{DETF}$.
This implies that the BAO2 data can give a fitting result with a slightly better accuracy.

\subsection{Various DE Diagnosis and Cosmic Age}

In this subsection, we study the impacts of different BAO data on the cosmic evolutions of various DE diagnosis (including $H(z)$, $q(z)$, $S^{(1)}_3(z)$ and $S^{(1)}_4(z)$) and the cosmic age $t(z)$.

\begin{figure*}
  \centering
  \resizebox{0.78\columnwidth}{!}{\includegraphics{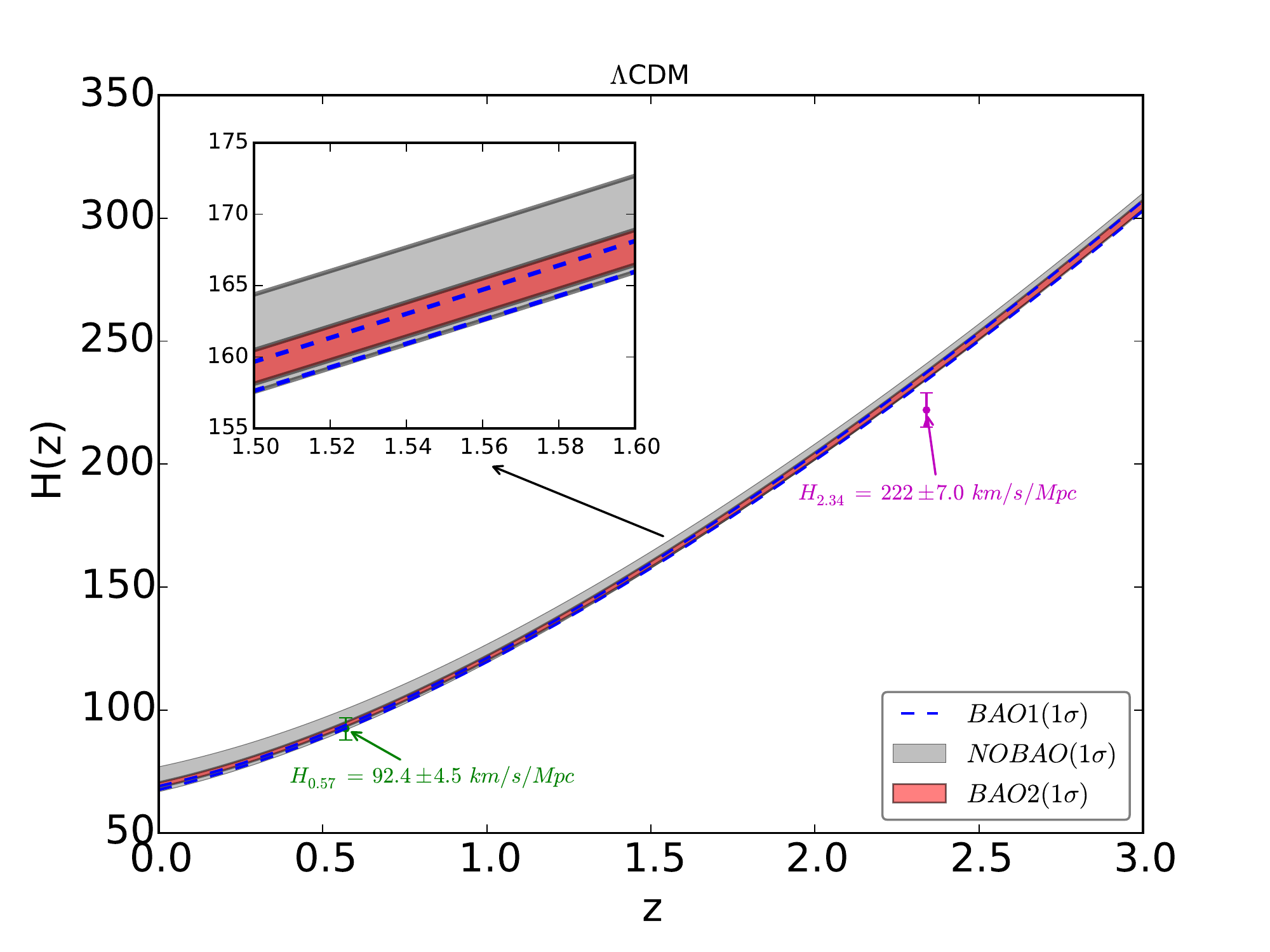}}
  \hspace{0.1\columnwidth}
  \resizebox{0.78\columnwidth}{!}{\includegraphics{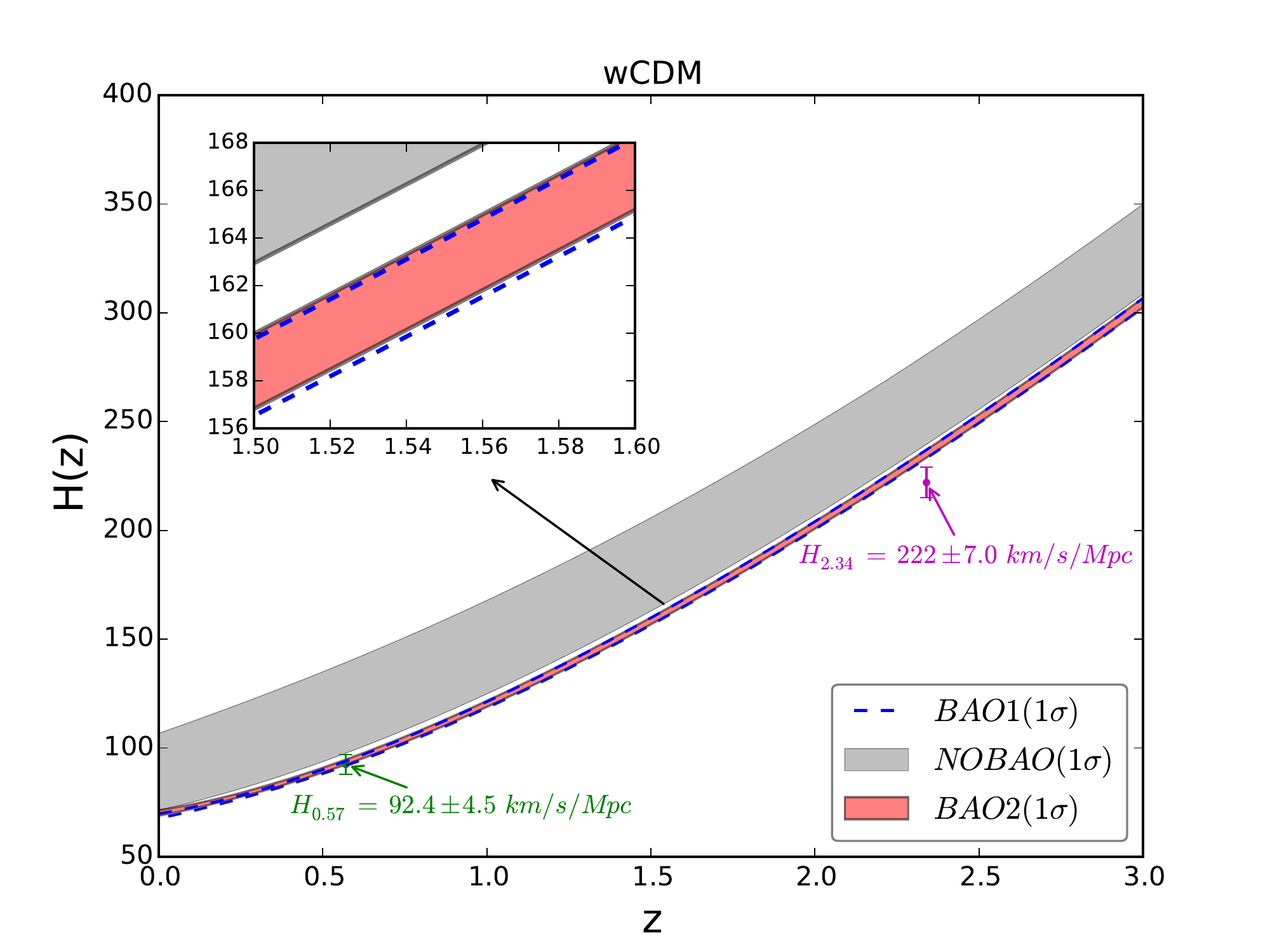}}
   \hspace{0.1\columnwidth}
  \resizebox{0.78\columnwidth}{!}{\includegraphics{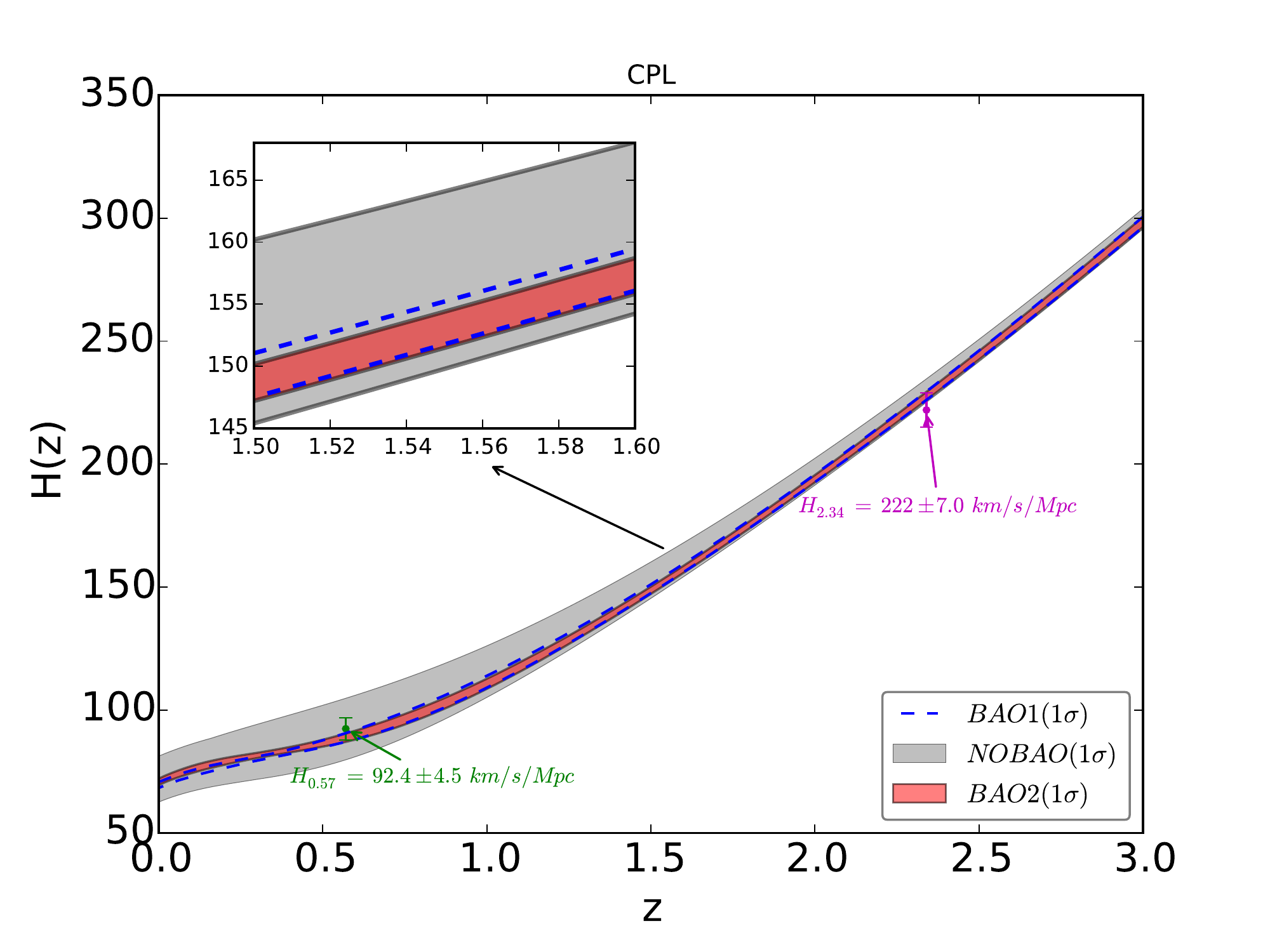}}
\caption{(color online). The 1$\sigma$ confidence regions of Hubble parameter $H(z)$ at redshift region $[0,3]$,
for the $\Lambda$CDM (upper left panel), the $w$CDM (upper right panel) and the CPL (lower panel) model,
where the data points of $H_{0.57}$ and $H_{2.34}$ are also marked by diamonds with error bars for comparison.
``BAO1'' (blue dashed lines), ``NO BAO'' (gray filled regions) and ``BAO2'' (red filled regions) denote the results given by the BAO1+SNLS3+Planck, the SNLS3+Planck and the BAO2+SNLS3+Planck data, respectively.
}
\label{fig:hz}
\end{figure*}

Fig.~\ref{fig:hz} shows the 1$\sigma$ confidence regions of Hubble parameter $H(z)$ at redshift region $[0,3]$ for the three DE models.
For comparison, two $H(z)$ data points, $H_{0.57}$ and $H_{2.34}$, are also marked by diamonds with error bars in this figure
\footnote{In fact, there actually exist a lot of other $H(z)$ measurements. see e.g. ~\citep{Ding2015}}.
It can be seen that for the $w$CDM model, the 1$\sigma$ confidence regions of $H(z)$ given by NO BAO seperate from the results given by BAO1 and BAO2.
This means that for the $w$CDM model, adding BAO data in the analysis will yield quite different $H(z)$ compared with the case without BAO data.
After adding BAO data in the analysis, we find that the data point $H_{0.57}$ can be easily accommodated in all the three DE models.
On the other side, the data point $H_{2.34}$ significantly deviates from the 1$\sigma$ regions of $H(z)$ in the $\Lambda$CDM and the $w$CDM model,
but it can be accommodated in the CPL model.
This means that the tension between $H_{2.34}$ and other cosmological observations~\citep{Sahni2014} can be reduced by considering the evolution of EoS $w$.
In addition, the 1$\sigma$ confidence regions of $H(z)$ given by different
BAO data are almost overlap, this means that using $H(z)$ diagram has great difficulty to distinguish the differences between the BAO1 and the BAO2 data.

\begin{figure*}
  \centering
  \resizebox{0.78\columnwidth}{!}{\includegraphics{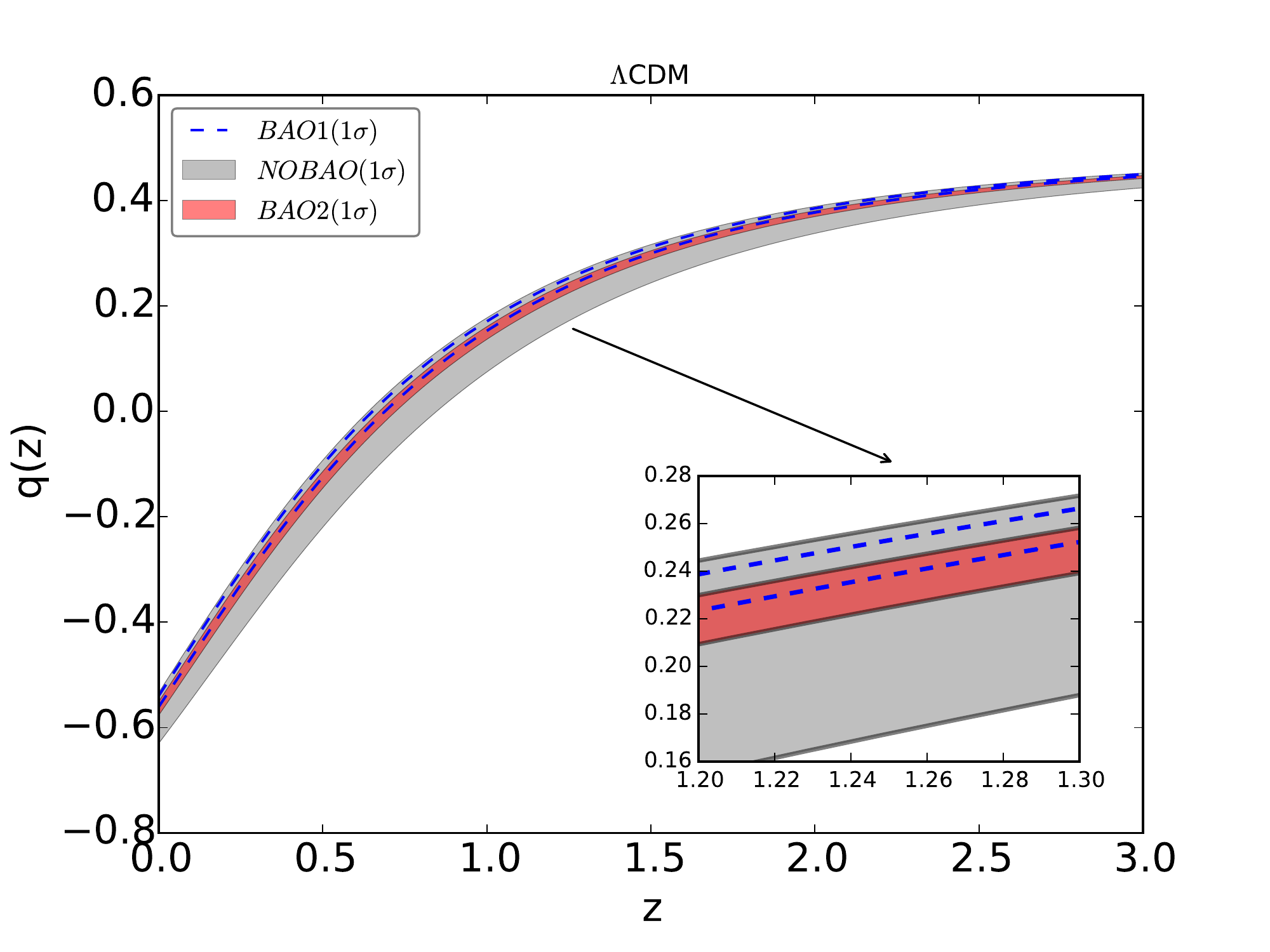}}
  \hspace{0.1\columnwidth}
  \resizebox{0.78\columnwidth}{!}{\includegraphics{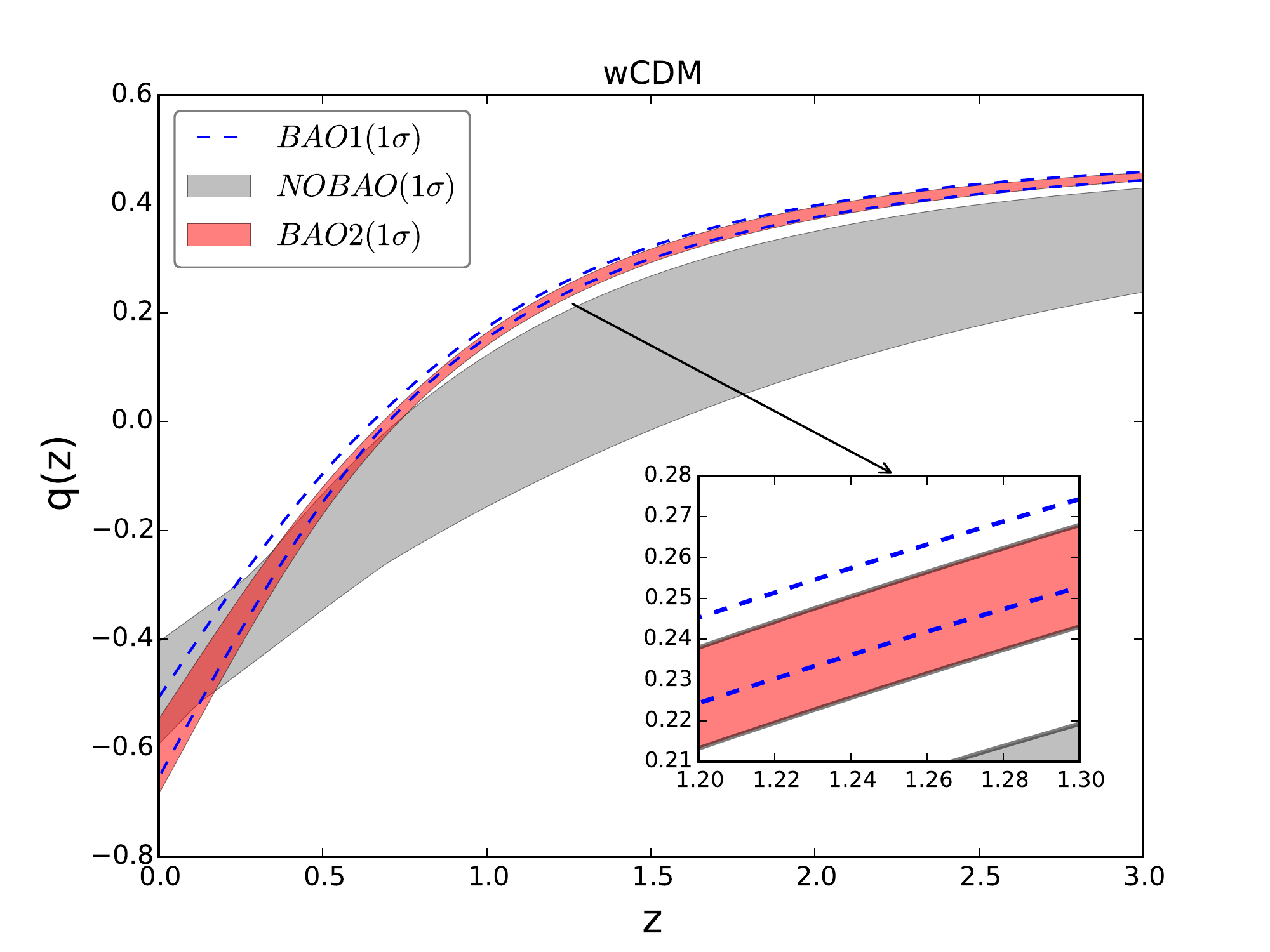}}
   \hspace{0.1\columnwidth}
  \resizebox{0.78\columnwidth}{!}{\includegraphics{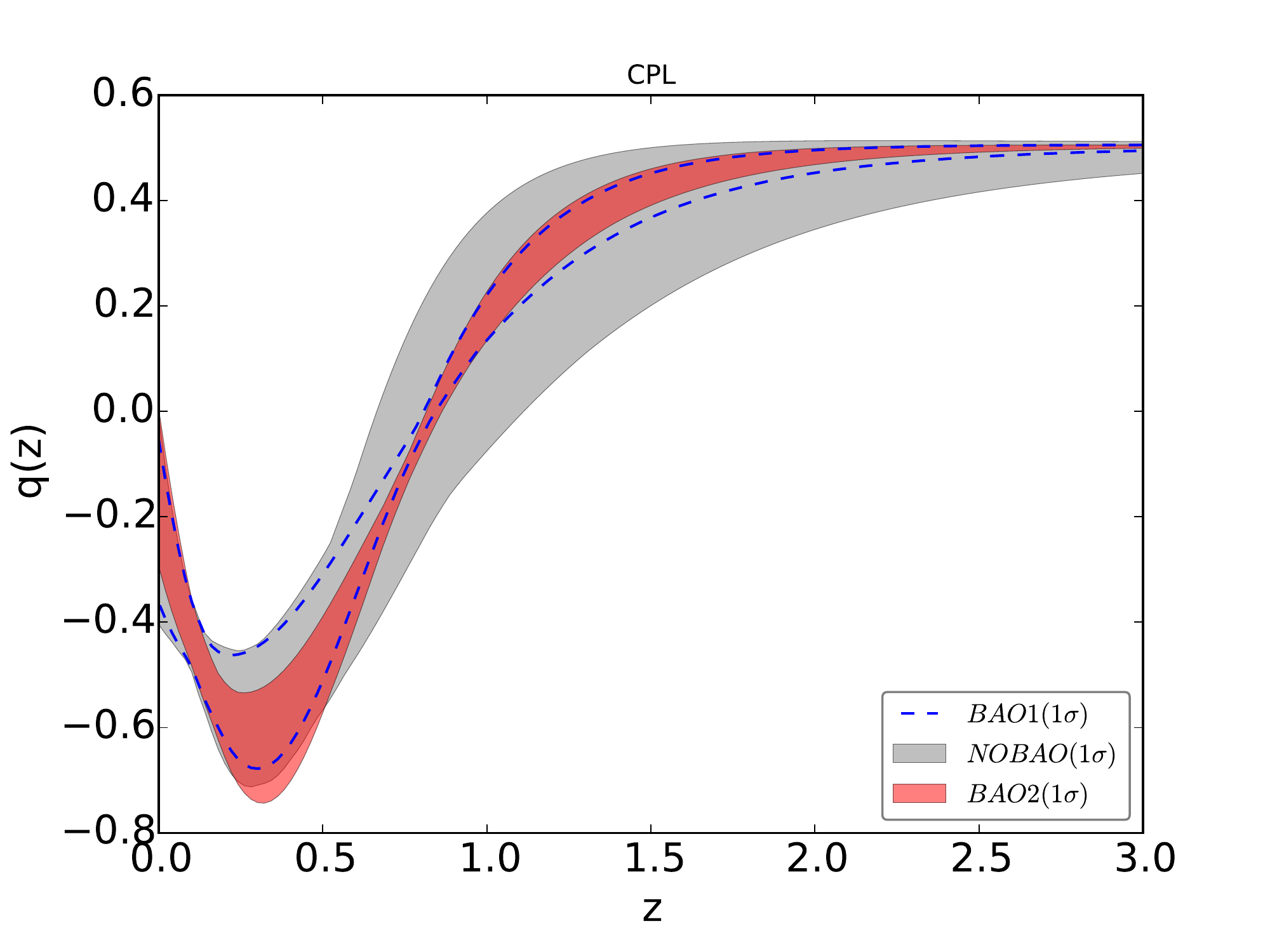}}
\caption{(color online). The 1$\sigma$ confidence regions of deceleration parameter $q(z)$ at redshift region $[0,3]$,
for the $\Lambda$CDM (upper left panel), the $w$CDM (upper right panel) and the CPL (lower panel) model.
``BAO1'' (blue dashed lines), ``NO BAO'' (gray filled regions) and ``BAO2'' (red filled regions) denote the results given by the BAO1+SNLS3+Planck, the SNLS3+Planck and the BAO2+SNLS3+Planck data, respectively.
}
\label{fig:qz}
\end{figure*}

We plot the 1$\sigma$ confidence regions of deceleration parameter $q(z)$ at redshift region $[0,3]$ in Fig.~\ref{fig:qz}.
From this figure we see that for the $w$CDM model, using BAO data will yield quite different $q(z)$ compared with the case without BAO data.
In addition, we find that for the CPL model, $q(z)$ achieves its minimum at $z \sim 0.3$, then starts to increase along with the decrease of redshift $z$.
This means that the current cosmic acceleration is probably slowing down.
In fact, the possibility of a slowing down cosmic acceleration has been proposed by Shafieloo et al. in~\citep{Shafieloo2009}; and in a recent work~\citep{YH2015}, we have proved that this extremly counterintuitive phenomenon is insensitive to the specific form of $w(z)$.
Moreover, for all the three models, we see that the
1$\sigma$ confidence regions of $q(z)$ given by different BAO data are almost
overlap, which implies that using $q(z)$ diagram still has difficulty to distinguish the differences between the BAO1 and the BAO2 data.

\begin{table*}
\centering
\caption{Deceleration-acceleration transition redshift $z_t$ for the $\Lambda$CDM, $w$CDM and CPL models, where the best-fit values are listed.
}
\label{tab:res_tranz}
\centering
\begin{tabular}{cccccccccccc}
\hline\hline &\multicolumn{2}{c}{$\Lambda$CDM}&&\multicolumn{2}{c}{$w$CDM}&&\multicolumn{2}{c}{CPL}\\
           \cline{2-3}\cline{5-6}\cline{8-9}
Quantity    & BAO1 & BAO2 && BAO1 & BAO2 && BAO1 & BAO2 \\ \hline
$z_t$              & $0.668$ 
                   & $0.696$&
                   & $0.676$
                   & $0.705$&
                   & $0.826$
                   & $0.836$\\ 
\hline
\end{tabular}
\end{table*}

In table \ref{tab:res_tranz}, we list the best fit values of deceleration-acceleration transition redshift $z_t$ for the three DE models.
From this table we see that, for all the cases, the transition redshifts are located at mediate redshift ($0.5<z_t<1$), which is consistent with the previous studies~\citep{Turner2002, Lima2012}.
Among the three DE models, the $\Lambda$CDM model corresponds to a smaller $z_t$, and the CPL model corresponds to a larger $z_t$.
In addition, for the two types of BAO data, BAO2 data always give a larger $z_t$.

\begin{figure*}
  \centering
  \resizebox{0.78\columnwidth}{!}{\includegraphics{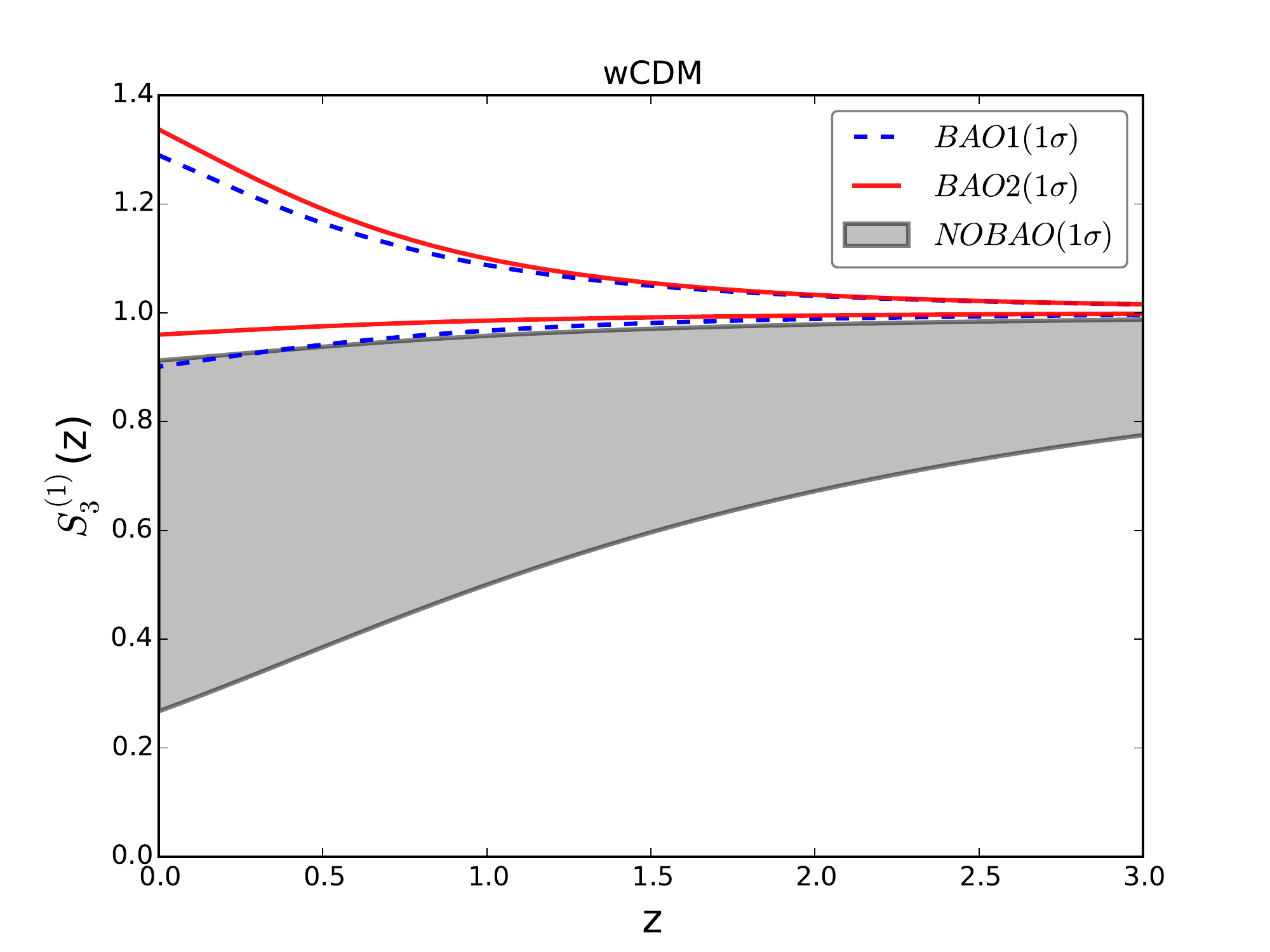}}
   \hspace{0.1\columnwidth}
  \resizebox{0.78\columnwidth}{!}{\includegraphics{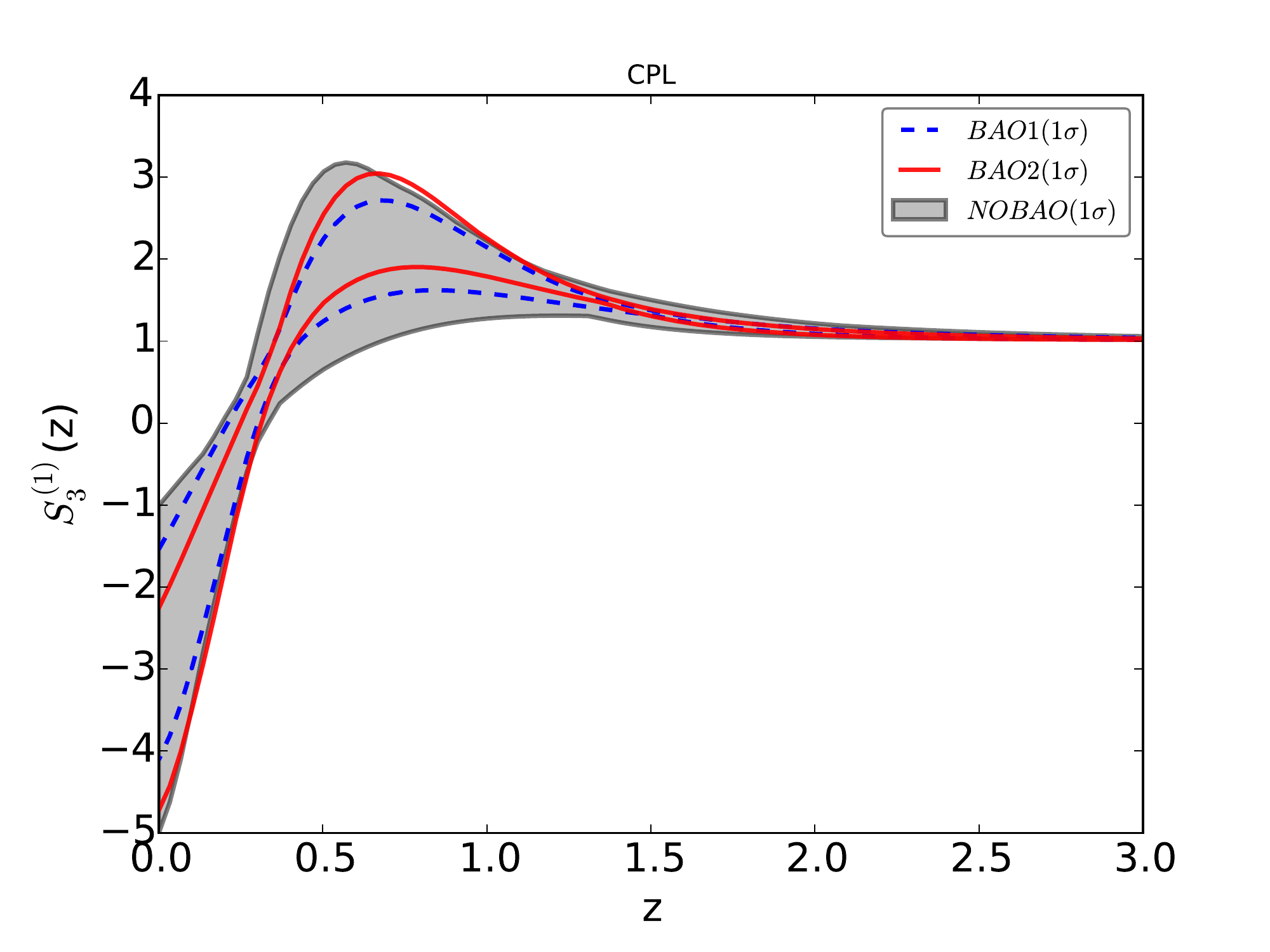}}
\caption{(color online). The 1$\sigma$ confidence regions of statefinder hierarchy $S^{(1)}_3(z)$ at redshift region $[0,3]$,
for the $w$CDM (left panel) and the CPL (right panel) model.
``BAO1'' (bule dashed line), ``BAO2'' (red solid line) and ``NO BAO'' (gray filled regions) denote the results given by the BAO1+SNLS3+Planck, the BAO2+SNLS3+Planck and the SNLS3+Planck data respectively.
}
\label{fig:statefinder3}
\end{figure*}

In Fig. \ref{fig:statefinder3}, we plot the 1$\sigma$ confidence regions of statefinder hierarchy
$S^{(1)}_3(z)$ at redshift region [0,3] for the $w$CDM model and the CPL model.
From this figure we see that  the evolution trajectory of
$S^{(1)}_3(z)$ given by the $w$CDM model is quite different from the result of the CPL model, which has a peak at $z\sim 0.6$.
This means that the statefinder hierarchy $S^{(1)}_3(z)$ is a powerful tool that has the ability to distinguish different DE models.
Moreover, we see that most of the 1$\sigma$ confidence regions of $S^{(1)}_3(z)$ given by the two BAO data are overlap.
This means that the statefinder $S^{(1)}_3(z)$ still does not have the ability to distinguish the effects of different BAO data.

\begin{figure*}
  \centering
  \resizebox{0.78\columnwidth}{!}{\includegraphics{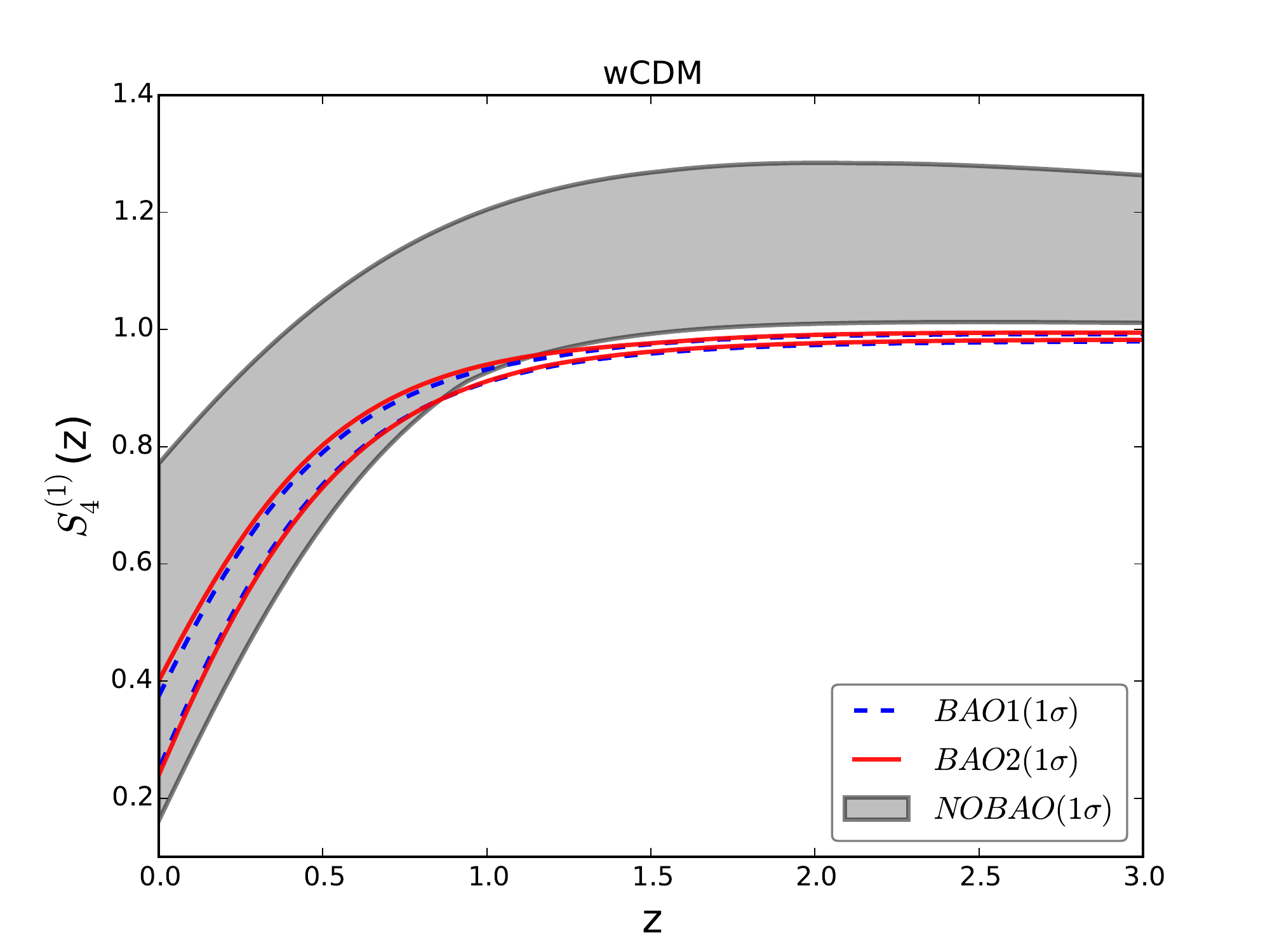}}
   \hspace{0.1\columnwidth}
  \resizebox{0.78\columnwidth}{!}{\includegraphics{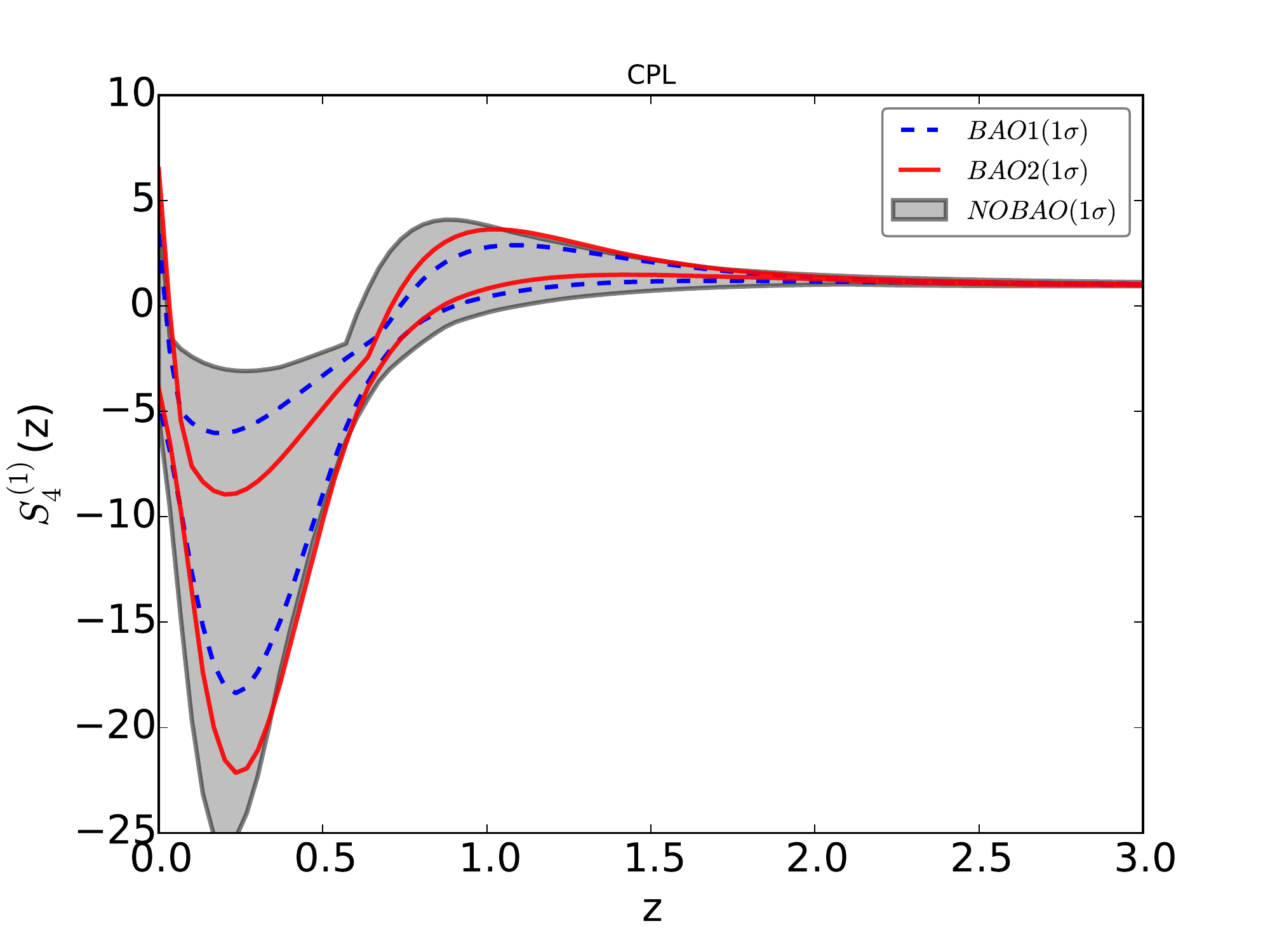}}
\caption{(color online). The 1$\sigma$ confidence regions of statefinder hierarchy $S^{(1)}_4(z)$ at redshift region $[0,3]$,
for the $w$CDM (left panel) and the CPL (right panel) model.
``BAO1'' (blue dashed line), ``BAO2'' (red solid line) and ``NO BAO'' (gray filled regions) denote the results given by the BAO1+SNLS3+Planck, the BAO2+SNLS3+Planck and the SNLS3+Planck data, respectively.
}
\label{fig:statefinder4}
\end{figure*}


In Fig. \ref{fig:statefinder4}, we plot the 1$\sigma$ confidence regions of statefinder hierarchy $S^{(1)}_4(z)$ at redshift region [0,3] for the $w$CDM model and the CPL model.
Again, we see that the evolution trajectory of
$S^{(1)}_4(z)$ given by the $w$CDM model is quite different from the results of the CPL model, which implies that the statefinder hierarchy $S^{(1)}_4(z)$ is also a powerful tool that has the ability to distinguish different DE models.
Moreover, we see that most of the 1$\sigma$ confidence regions of $S^{(1)}_4(z)$ given by the two BAO data are overlap.
This means that the effects of different BAO data can not be distinguishthed by using the statefinder $S^{(1)}_4(z)$ either.

In addition, from Figs. \ref{fig:statefinder3} and \ref{fig:statefinder4} we see that for the $w$CDM model, the 1$\sigma$ confidence regions of $S^{(1)}_3(z)$ given by BAO1 and BAO2 seperate from the 1$\sigma$ regions given by NO BAO; in contrast, the 1$\sigma$ regions of $S^{(1)}_4(z)$ given by BAO1 and BAO2 are enclosed by the 1$\sigma$ regions of NO BAO at $z<1$.
This means that compared with $S^{(1)}_4(z)$, $S^{(1)}_3(z)$ is a better diagnosis tool (for similar results, see~\citep{Myrzakulov2013etc}).

\begin{figure*}
  \centering
  \resizebox{0.78\columnwidth}{!}{\includegraphics{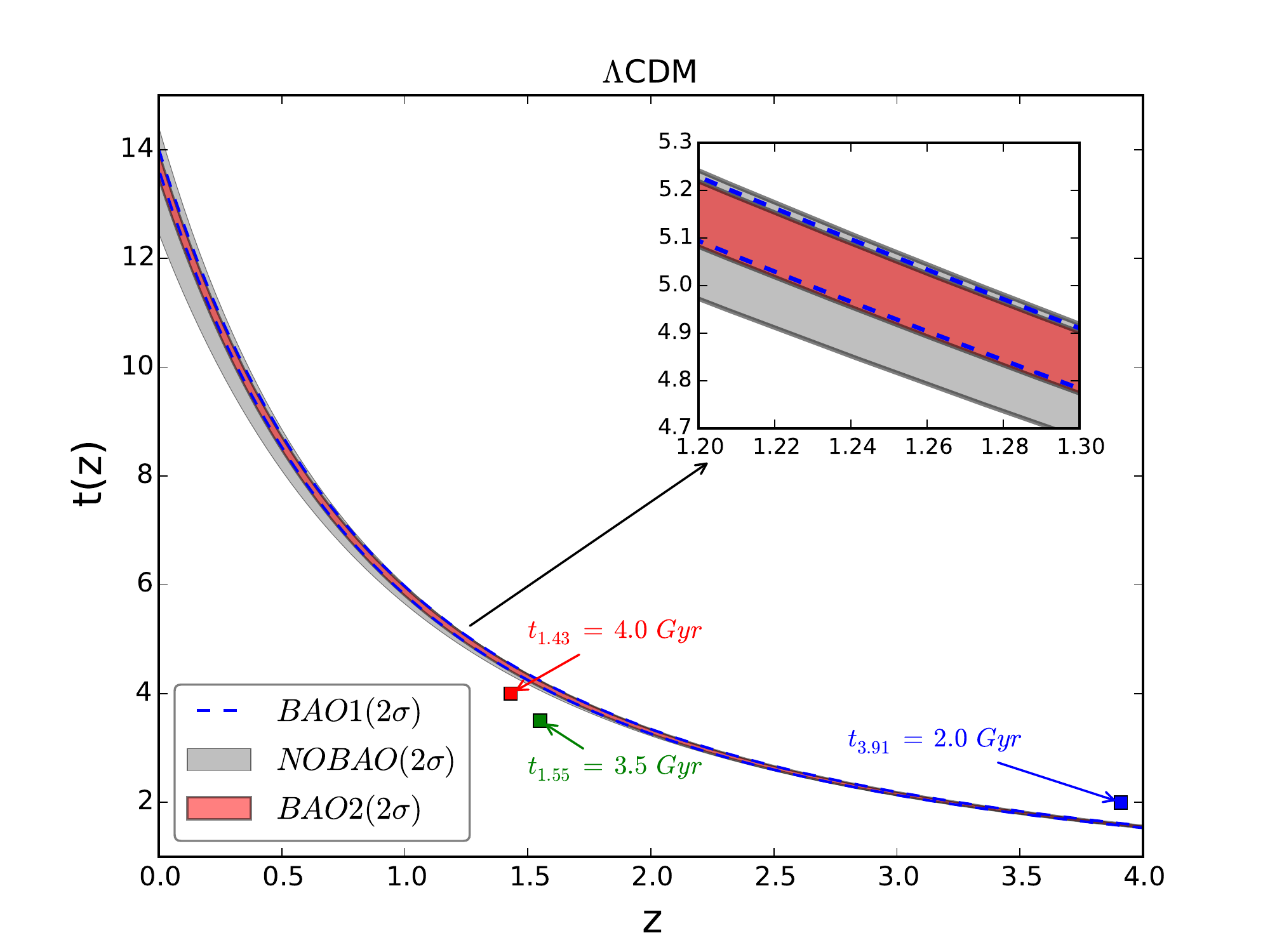}}
  \hspace{0.1\columnwidth}
  \resizebox{0.78\columnwidth}{!}{\includegraphics{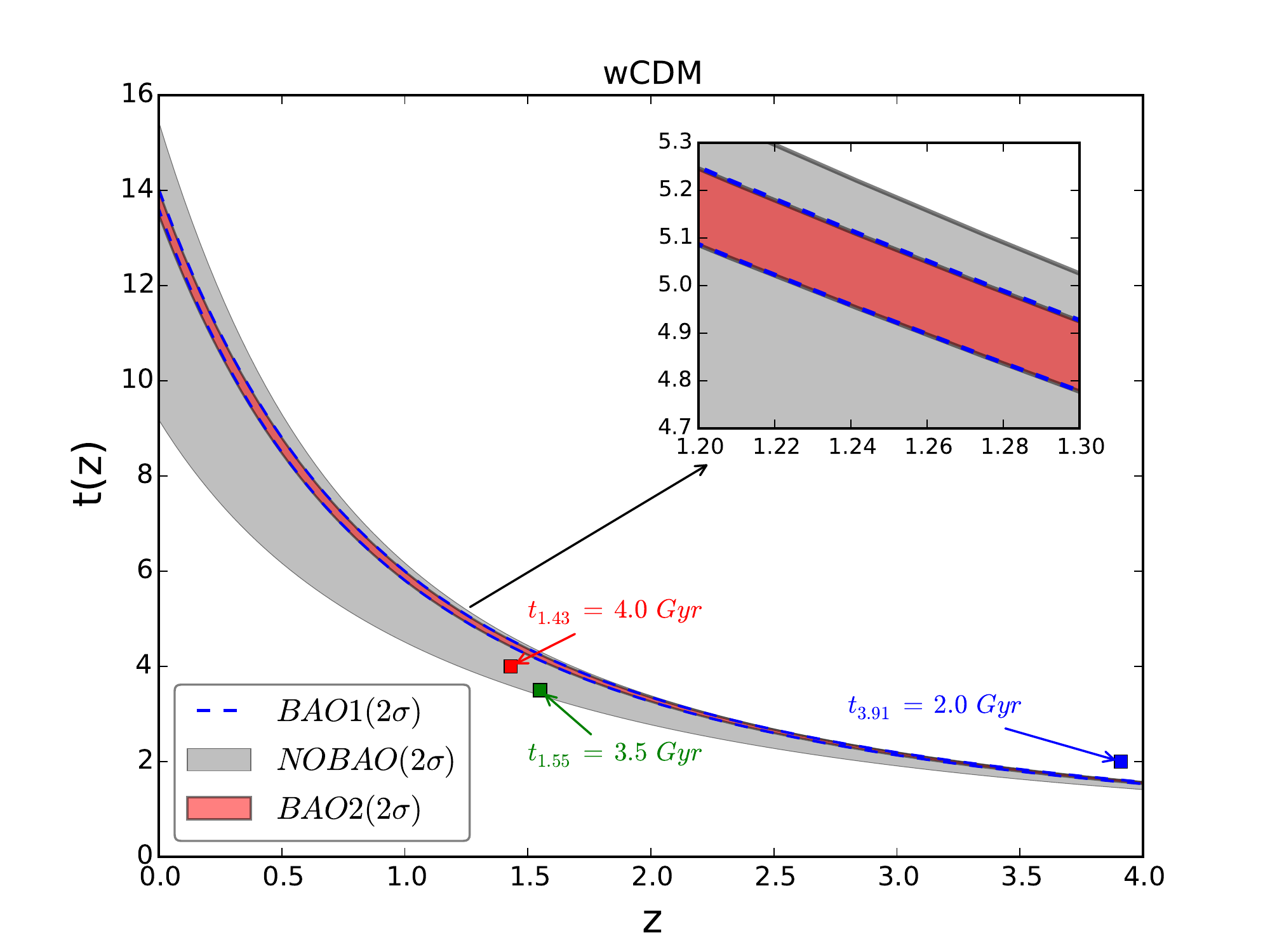}}
   \hspace{0.1\columnwidth}
  \resizebox{0.78\columnwidth}{!}{\includegraphics{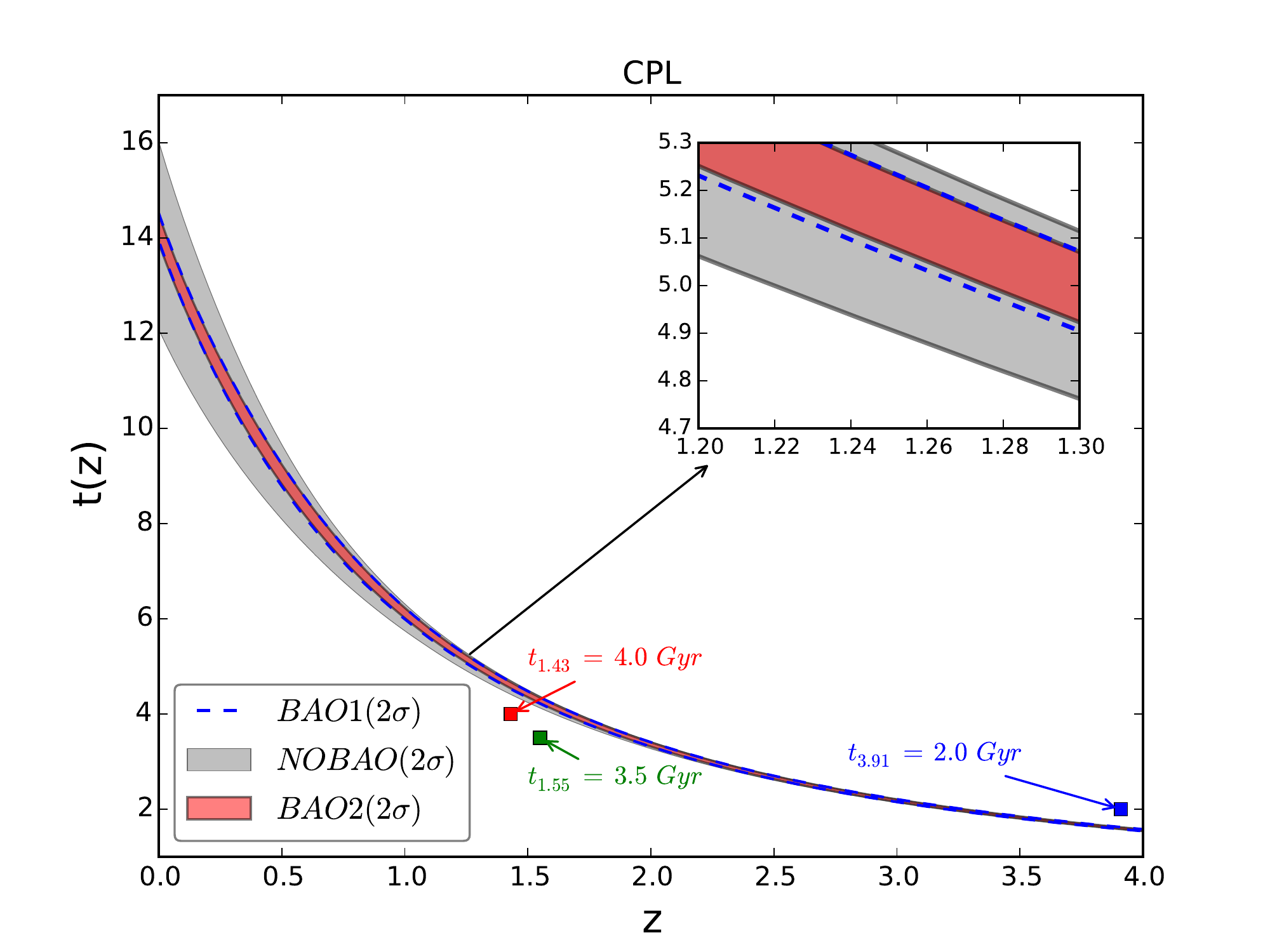}}
  \caption{(color online). The 2$\sigma$ confidence regions of cosmic age $t(z)$ at redshift region $[0,4]$,
for the $\Lambda$CDM (upper left panel), the $w$CDM (upper right panel) and the CPL (lower panel) model.
``BAO1'' (blue dashed lines), ``NO BAO'' (gray filled regions) and ``BAO2'' (red filled regions) denote
the results given by the BAO1+SNLS3+Planck, the SNLS3+Planck and the BAO2+SNLS3+Planck data, respectively.
}
\label{fig:tz}
\end{figure*}

In Fig. \ref{fig:tz}, we plot the 2$\sigma$ confidence regions of cosmic age $t(z)$ at redshift
region [0,4] for the three DE models, where the three $t(z)$ data points, $t_{1.43}$, $t_{1.55}$ and $t_{3.91}$,
are also marked by Squares for comparison.
We find that both $t_{1.43}$ and $t_{1.55}$ can be easily accommodated in all the three DE models,
but the position of $t_{3.91}$ is significantly higher than the 2$\sigma$ upper bounds of all the three DE models.
This means that the existence of the old quasar APM 08279+5255 can not be explained in the standard cosmology.
This result is consistent with the conclusions of pervious studies \cite{Alcaniz03}.
In addition, the 2$\sigma$ regions of $t(z)$ given by BAO1 and BAO2 are almost overlap,
showing that the impacts of different BAO data can not be distinguished by using the age data of OHRO.

\section{Summary}
\label{sec:conclusion}

In this work, we explore the cosmological implications of two types of BAO data. These two types BAO data are obtained by using the spherically averaged 1D galaxy clustering statistics (BAO1 data) and the anisotropic 2D galaxy clustering statistics (BAO2 data), respectively.
So far as we know,
the effects of different BAO data on cosmology-fits and corresponding cosmological consquences have not been studied in the past.
So the main aim of our work is presenting a comprehensive and systematic investigation on the cosmological implications of different BAO data.
To make a comparison, we also take into account the case without any BAO data.

Making use of the BAO1 and the BAO2 data,
as well as the SNLS3 SNe Ia sample and the Planck distance priors data,
we give the cosmological constraints of the $\Lambda$CDM, the $w$CDM and the CPL model.
Then, according to the cosmological fitting results,
we study the impacts of different BAO data on cosmological consquences, including parameter space, EoS, FoM, deceleration-acceleration transition redshift, Hubble parameter $H(z)$, deceleration parameter $q(z)$,
statefinder hierarchy $S^{(1)}_3(z)$ and $S^{(1)}_4(z)$, and cosmic age $t(z)$.

We find that: (1) For all the three DE models, NO BAO data always give a smallest fractional matter density $\Omega_{m0}$, a largest fractional curvature density $\Omega_{k0}$ and a largest Hubble constant $h$; in contrast, BAO1 data always give a largest $\Omega_{m0}$, a smallest $\Omega_{k0}$ and a smallest $h$ (see table~\ref{tab:res_cos}).
(2) For the $w$CDM and the CPL model, NO BAO data always give a largest $w$; in contrast, BAO2 always give a smallest $w$ (see Figs.~\ref{fig:wcdm_w} and ~\ref{fig:cpl_wz}).
(3) Compared with the case of BAO1, BAO2 data always give a slightly larger FoM, and thus can give a cosmological constraint with a slightly
better accuracy (see table~\ref{tab:res_fomr}).
(4) The impacts of different BAO data on the cosmic evolution and the comic age are very small, and can not be distinguished by using various dark energy diagnosis (see Figs.~\ref{fig:hz}, ~\ref{fig:qz}, ~\ref{fig:statefinder3} and \ref{fig:statefinder4}) and the cosmic age data (see Figs.~\ref{fig:tz}).

It would be interesting to further explore the cosmological implications of these two types of BAO data by considering some other factors, such as interaction between dark sectors \citep{Guo07,LiYH13}, sterile neutrinos \citep{Zhang2014a}, and cosmic fate \citep{Caldwell03}.
These issues will be studied in future works.

\begin{acknowledgements}

SW is supported by the National Natural Science Foundation of China under Grant No. 11405024
and the Fundamental Research Funds for the Central Universities under Grant No. 16lgpy50.
ML is supported by the National Natural Science Foundation of China (Grant No. 11275247, and Grant No. 11335012)
and a 985 grant at Sun Yat-Sen University.

\end{acknowledgements}

\end{document}